\documentclass[preprint,pra]{revtex4}%
\usepackage{amsfonts}
\usepackage{amsmath}
\usepackage{amssymb}
\usepackage{graphicx}%
\providecommand{\U}[1]{\protect\rule{.1in}{.1in}}

\ifx\pdfoutput\relax\let\pdfoutput=\undefined\fi
\newcount\msipdfoutput
\ifx\pdfoutput\undefined\else
\ifcase\pdfoutput\else
\ifx\paperwidth\undefined\else
\ifdim\paperheight=0pt\relax\else\pdfpageheight\paperheight\fi
\ifdim\paperwidth=0pt\relax\else\pdfpagewidth\paperwidth\fi
\fi\fi\fi
\begin{document}
\title{Properties of the Anomalous States of Positronium}
\author{Chris W. Patterson}
\affiliation{Centre for Quantum Dynamics, Griffith University, Nathan QLD 4111, Australia}
\altaffiliation{Guest Scientist, Theoretical Division, Los Alamos National Laboratory, NM
87545 USA}

\begin{abstract}
\textit{It is shown that there are anomalous bound-state solutions to the
two-body Dirac equation for an electron and a positron interacting via an
electromagnetic potential. These anomalous solutions have quantized
coordinates at nuclear distances (fermi) and are orthogonal to the usual
atomic positronium bound-states as shown by a simple extension of the
Bethe-Salpeter equation. It is shown that the anomalous states have many
properties which correspond to those of neutrinos.}

\end{abstract}
\maketitle

\section{Introduction}

As shown in previous papers \cite{Scott1992} , \cite{Patterson2019}, both the
atomic states $\Psi_{Atom}$ and anomalous states $\Psi_{An}$ of positronium
admit bound-state solutions to the two-body Dirac equation (TBDE),%
\begin{equation}
H\Psi=(H_{0}+\Phi(\rho)\Psi=E\Psi, \label{1.1}%
\end{equation}
with an instantaneous Coulomb potential $\Phi(\rho)=$ $\Phi_{C}=-e^{2}/\rho$,
where $H_{0}$ is the free-particle Hamiltonian in the rest frame. By
definition, the anomalous states $\Psi_{An}$ are solutions to (\ref{1.1}) with
$\Phi(\rho)=0$ such that%
\begin{equation}
H_{0}\Psi_{An}=0\mathbf{.} \label{1.2}%
\end{equation}
Let the anomalous bound-states $\Psi_{DV}$ be linear combinations of
$\Psi_{An}$ which are solutions to (\ref{1.1}). From (\ref{1.1}) and
(\ref{1.2}), the $\Psi_{DV}$ are then bound-state solutions to the simple
equation
\begin{equation}
\Phi_{C}(\rho)\Psi_{DV}=E\Psi_{DV}. \label{1.3}%
\end{equation}
It was shown in \cite{Patterson2019} that the bound-state solutions to
(\ref{1.3}) $\Psi_{DV}$ can be found by using the discrete variable (DV)
representation \cite{Light2000}. That is, the solutions to (\ref{1.3}) are the
DV states $\Psi_{DV}$ in which the coordinate $\rho$ is quantized at discrete
separations or $\Psi_{DV}=G\delta(\rho-\rho_{i})$ with $\rho_{i}$ at $fermi$
distances, where the factor $G$ includes the dependence on the other
coordinates as well as the spinors. Such DV states can be found by
diagonalizing $\Phi_{C}$ in the $\Psi_{An}$ bases only if the $\Psi_{An}$
bases is a complete set in the $\rho$ coordinate. The use of the word
`discrete' in DV theory refers to the quantization of the radial coordinate
$\rho=\rho_{i}$, in contrast to the normal quantization of momentum for
free-particles. Note that the free-particle anomalous states $\Psi_{An}$ in
(\ref{1.2}) are distinguished from the bound-states $\Psi_{DV}$ of (\ref{1.3})
which are comprised of these free-particle states. This distinction has been
made because not all anomalous states $\Psi_{An}$ can form a complete set in
$\rho$ and can thereby admit bound-states solutions $\Psi_{DV}$ of (\ref{1.1})
or (\ref{1.3}) using DV theory.

The anomalous bound-states $\Psi_{DV}$, with total angular momentum $J=0$,
were shown in \cite{Patterson2019} to have properties which were quite
distinct from their atomic counterparts. Besides being bound at nuclear
distances $\rho_{i}$, it was found that these DV states are dark and stable.
That is, they cannot absorb or emit light, nor can the electron and positron
annihilate or dissociate. This unusual behavior was explained by solving both
the TBDE and the Bethe-Salpeter equation (BSE) \cite{Salpeter1951},
\cite{Salpeter1952} with a Coulomb potential. It was found that these DV
states $\Psi_{DV}$ form doublets with spin $S_{z}=0$ and energy $E_{i}%
=-e^{2}/\rho_{i}$

However, in \cite{Patterson2019}, it was also found that the Coulomb potential
in the TBDE (\ref{1.1}) erroneously mixes the anomalous bound-states with the
atomic bound-states, resulting in the unusual behavior of the atomic
ground-state wavefunction near the origin as found in \cite{Scott1992}. This
erroneous mixing, while small, was shown not to occur at all for the
relativistically correct BSE because of the different time propagation for
anomalous bound-states and atomic bound-states. In particular, it was shown,
using the BSE, that the time dependence of the atomic states was determined by
the two-body Feynman propagator $K_{F}^{2}$, whereas the time dependence of
the anomalous states was determined by the two-body retarded propagator
$K_{R}^{2}$. As a result, the anomalous bound-states for the instantaneous
Coulomb potential were temporally orthogonal to the atomic bound-states and
could not be mixed.

It was further argued in \cite{Patterson2019} that the anomalous bound-states
were, themselves, both mathematically viable and necessary for completeness in
space-time when using the BSE. It seems appropriate that their properties
should be investigated further to see if they could exist physically. In
\cite{Patterson2019}, only the DV states with total angular momentum $J=0$
were considered in order to show their influence on the atomic $J=0$
wavefunctions and energies when using the TBDE. In this paper, the DV states
for all $J$ are considered with the restriction to $J_{z}=0$. It is found that
there are then four different possible DV states corresponding to a $S=0$
doublet and a $S=1$ doublet for each $J>0$. Like the $J=0$ DV states, these DV
states for $J\neq0$ are also dark and stable. Including the magnetic potential
$\Phi_{M}$, the $S=0$ DV doublet has energy $E_{i}^{0}=2e^{2}/\rho_{i}$ and
the $S=1$ DV doublet has energy $E_{i}^{1}=0$.

It is also shown that the Lorentz boost reduces the symmetry from spherical
$(\rho,\theta)$ to cylindrical $(r,z)$ where $\widehat{z}$ is in the direction
of motion. Because of this dynamical symmetry breaking, in the moving frame
the DV doublet states with $S_{z}=\pm1$ can only occur in the $z=0$ plane with
$\theta=\pi/2$, so that they are oriented perpendicular to the direction of
motion. On the other hand the DV doublet states with $S_{z}=0$ can occur for
any $\theta$. Remarkably, using a Lorentz boost for the DV states, it is
proven that doublets with $\theta=\pi/2$ transform like Majorana fermions
instead of bosons. It is then shown that the $S_{z}=\pm1$ fermions with mass
$M_{i}^{1}=0$ have well defined chirality and helicity as expected for a zero
mass fermions.

Finally, it is shown that all of the unusual properties of the anomalous
bound-states $\Psi_{DV}$ are a result of the fact that either the electron or
the positron (but not both) must be in a negative energy state. As a result,
one's normal understanding of quantum and classical mechanics can be
misleading. However, it is shown that these anomalous bound-states still have
a simple classical correspondence which aids in understanding their novel properties.

In order to make this paper reasonably self-contained, the notation and the
salient developments of the previous work are briefly reviewed below so that
the reader can better understand the three equations above. In this review,
comparisons are made between the TBDE and the BSE which are especially
important to the understanding of the anomalous states.

(Note that the natural units $c=1$ and $\hbar=1$ are used below except for
cases where clarity is needed.)

\section{Review}

The Hamiltonian of the TBDE for a free electron and positron, $H_{0}$, in
(\ref{1.1}) in the moving frame is the sum of the individual Dirac
Hamiltonians,
\begin{equation}
H_{0}\Psi=\{\boldsymbol{\alpha}_{e}\cdot\boldsymbol{p}_{e}+\boldsymbol{\alpha
}_{p}\cdot\boldsymbol{p}_{p}\text{\textbf{\ }}+\text{\textbf{\ }}(\gamma
_{e4}+\gamma_{p4})m\}\Psi=E\Psi. \label{2.1}%
\end{equation}
Transforming this equation using relative coordinates and their conjugate
momenta,%
\begin{align}
\boldsymbol{\rho}  &  =(\mathbf{r}_{e}-\mathbf{r}_{p}),\ \ \boldsymbol{\pi
}=\mathbf{(p}_{e}-\mathbf{p}_{p})/2,\ \label{2.2}\\
\mathbf{R}  &  =(\mathbf{r}_{e}+\mathbf{r}_{p})/2,\ \ \ \mathbf{P}^{\prime
}=\mathbf{p}_{e}+\mathbf{p}_{p},\nonumber
\end{align}
one finds \cite{Scott1992}, \cite{Malenfant1988},%

\begin{align}
H_{0}\Psi &  =\{K+\mathcal{M}+\frac{(\boldsymbol{\alpha}_{e}%
+\boldsymbol{\alpha}_{p})}{2}\cdot\mathbf{P^{\prime}}\}\Psi=E\Psi
,\label{2.4}\\
K  &  =(\boldsymbol{\alpha}_{e}-\boldsymbol{\alpha}_{p})\cdot\boldsymbol{\pi
}\mathbf{,}\nonumber\\
\mathcal{M}  &  =m(\gamma_{e4}+\gamma_{p4}),\nonumber
\end{align}
where $K$ is the two-body kinetic operator and $\mathcal{M}$ is the two-body
mass operator. The prime is used for $\mathbf{P^{\prime}}$ to indicate the
frame is moving rather than at rest. The matrix elements of $\alpha_{k}$ and
$\gamma_{4}$ in the bases of the two Dirac-spinors $\mathbf{e}_{1}$ and
$\mathbf{e}_{2}$ are given by
\begin{align}
\alpha_{k}  &  =i\gamma_{4}\gamma_{k}~=\sigma_{k}\left(
\begin{array}
[c]{cc}%
0 & 1\\
1 & 0
\end{array}
\right)  ,\ (for\ k=1,2,3),\\
\gamma_{4}  &  =\left(
\begin{array}
[c]{cc}%
1 & 0\\
0 & -1
\end{array}
\right)  ,\nonumber
\end{align}
so that $\alpha_{k}\mathbf{e}_{1}=\sigma_{k}\mathbf{e}_{2},$ $\alpha
_{k}\mathbf{e}_{2}=\sigma_{k}\mathbf{e}_{1},$ $\gamma_{4}\mathbf{e}%
_{1}=\mathbf{e}_{1},$ and $\gamma_{4}\mathbf{e}_{2}=-\mathbf{e}_{2}$. One can
then define the four Dirac-spinors $\mathbf{e}_{ij}=\mathbf{e}_{i}%
\times\mathbf{e}_{j}$ for the direct product wavefunctions $\Psi=\psi
(e)\times\psi(p)$ such that%
\begin{align}
\Psi &  \equiv\binom{_{\psi_{1}(e)\psi_{2}(p)}^{\psi_{1}(e)\psi_{1}(p)}%
}{_{\psi_{2}(e)\psi_{2}(p)}^{\psi_{2}(e)\psi_{1}(p)}}\equiv\binom{_{\Psi_{12}%
}^{\Psi_{11}}}{_{\Psi_{22}}^{\Psi_{21}}},\label{2.8}\\
&  \equiv\Psi_{11}\mathbf{e}_{11}+\Psi_{12}\mathbf{e}_{12}+\Psi_{21}%
\mathbf{e}_{21}+\Psi_{22}\mathbf{e}_{22}.\nonumber
\end{align}
With this notation, one finds the simple equations for the mass operators
$\gamma_{e4}$ and $\gamma_{p4}$ of $\mathcal{M}$,%
\begin{align*}
\gamma_{e4}\Psi &  =(\Psi_{11}\mathbf{e}_{11}+\Psi_{12}\mathbf{e}_{12}%
-\Psi_{21}\mathbf{e}_{21}-\Psi_{22}\mathbf{e}_{22}),\\
\gamma_{p4}\Psi &  =(\Psi_{11}\mathbf{e}_{11}-\Psi_{12}\mathbf{e}_{12}%
+\Psi_{21}\mathbf{e}_{21}-\Psi_{22}\mathbf{e}_{22}).
\end{align*}
It is useful to use a bases where $\Psi_{g}$ is symmetric and $\Psi_{u}$ is
antisymmetric under the simultaneous exchange $\mathbf{e}_{11}\leftrightarrow
\mathbf{e}_{22}$ and $\mathbf{e}_{12}\leftrightarrow\mathbf{e}_{21}$, such
that
\begin{align}
\Psi &  =\Psi_{g}+\Psi_{u},\label{2.10}\\
\Psi_{g}  &  =%
\frac12
\{(\Psi_{11}+\Psi_{22})(\mathbf{e}_{11}+\mathbf{e}_{22})+(\Psi_{12}+\Psi
_{21})(\mathbf{e}_{12}+\mathbf{e}_{21})\},\nonumber\\
\Psi_{u}  &  =%
\frac12
\{(\Psi_{11}-\Psi_{22})(\mathbf{e}_{11}-\mathbf{e}_{22})+(\Psi_{12}-\Psi
_{21})(\mathbf{e}_{12}-\mathbf{e}_{21})\}.\nonumber
\end{align}
The $\Psi_{g}$ and $\Psi_{u}$ states are only coupled by the mass operator
$\mathcal{M}$ in (\ref{2.4}) where%
\begin{equation}
\mathcal{M}(\mathbf{e}_{11}\pm\mathbf{e}_{22})=2m(\mathbf{e}_{11}\mp
\mathbf{e}_{22}). \label{2.10a}%
\end{equation}

One can also define the four Pauli-spinors $\chi_{\pm%
\frac12
}(e)\chi_{\pm%
\frac12
}(p)$ in terms of the states $\Omega_{S_{z}}^{S}$ with total spin $S$ and its
$z$ component $S_{z}$, so that
\begin{align}
\Omega_{0}^{0}  &  =\sqrt{%
\frac12
}[\chi_{%
\frac12
}(e)\chi_{-%
\frac12
}(p)-\chi_{-%
\frac12
}(e)\chi_{%
\frac12
}(p)],\label{2.9}\\
\Omega_{-1}^{1}  &  =\chi_{-%
\frac12
}(e)\chi_{-%
\frac12
}(p),\nonumber\\
\Omega_{0}^{1}  &  =\sqrt{%
\frac12
}[\chi_{%
\frac12
}(e)\chi_{-%
\frac12
}(p)+\chi_{-%
\frac12
}(e)\chi_{%
\frac12
}(p)],\nonumber\\
\Omega_{1}^{1}  &  =\chi_{%
\frac12
}(e)\chi_{%
\frac12
}(p).\nonumber
\end{align}
The singlet state $\Omega_{0}^{0}$ is antisymmetric $(X=-1)$ and triplet
states $(\Omega_{-1}^{1},\Omega_{-0}^{1},\Omega_{1}^{1})$ are symmetric
$(X=1)$ under particle exchange $\mathbf{X}$, where $\mathbf{X}\Omega_{S_{z}%
}^{S}=X\Omega_{S_{z}}^{S}$. The Dirac wavefunctions $\Psi$ include terms of
the sixteen possible spinors $e_{ij}\Omega_{S_{z}}^{S}$ for the four
Dirac-spinors $e_{ij}$ and the four Pauli-spinors $\Omega_{S_{z}}^{S}$.

The free-particle solutions of (\ref{2.4}) are direct products $\Psi_{\pm\pm
}=\psi_{\pm}(e)\times\psi_{\pm}(p)$ of the single-particle solutions for
positive and negative energy states $\psi_{+}$ and $\psi_{-}$ with energies,%
\[
E_{\pm\pm}=e_{e}+e_{p}=\pm\sqrt{p_{e}^{2}+m^{2}}\pm\sqrt{p_{p}^{2}+m^{2}}.
\]

Now let \textbf{$P^{\prime}$}$\mathbf{=}\mathbf{0}$ in (\ref{2.4}) so that
$H_{0}=K+\mathcal{M}$. Using (\ref{2.2}), one finds that $\mathbf{p}%
_{e}=-\mathbf{p}_{p}$ and $p_{e}^{2}=p_{p}^{2}=\pi^{2}$ where $\mathbf{\pi}$
is the relative momentum. The energies of the free-particle states are then
\begin{align}
E_{\pm\pm}  &  =\pm e\pm e,\label{2.5}\\
e  &  =\sqrt{\pi^{2}+m^{2}}.\nonumber
\end{align}
When using the BSE for \textbf{$P^{\prime}$}$\mathbf{=}\mathbf{0}$, it is
useful to divide the free -particle solutions into either atomic\textbf{
}states or anomalous\textbf{ }states. The atomic free-particle states\ $\Psi
_{Atom}$ have wavefunctions $\Psi_{++}$ and $\Psi_{--}$ \cite{Salpeter1951},
\cite{Salpeter1952} with energies
\begin{equation}
E_{++}=2e,~\ E_{--}=-2e. \label{2.6}%
\end{equation}
The anomalous free -particle states $\Psi_{An}$ have wavefunctions $\Psi_{+-}$
and $\Psi_{-+}$ \cite{Patterson2019} with energies as in (\ref{1.2})%
\begin{equation}
E_{+-}=0,~\ E_{-+}=0. \label{2.7}%
\end{equation}

Both the atomic bound-states and anomalous bound-states are solutions to the
TBDE (\ref{1.1}) for $\Phi=\Phi_{C}$. In the momentum representation with a
Coulomb potential, one can write (\ref{1.1}) (often called the Breit equation)
as the integral equation,%

\begin{equation}
H_{0}\Psi(\boldsymbol{\pi})-\frac{e^{2}}{2\pi^{2}}\int d^{3}k\frac
{1}{\boldsymbol{k\cdot k}}\Psi(\boldsymbol{k+\pi})=E\Psi(\boldsymbol{\pi}).
\label{2.11}%
\end{equation}
Note that this equation allows the $\Psi_{Atom}$ and $\Psi_{An}$ states to be
mixed by the Coulomb potential. As shown in \cite{Patterson2019}, the two
different bound-state solutions to (\ref{1.1}) or (\ref{2.11}), $\Psi_{Atom}$
and $\Psi_{DV}$, have different time behaviors. Together, these two solutions
form a complete orthonormal set in space-time. The $\Psi_{Atom}$ and
$\Psi_{DV}$ correspond to solutions to two different BSEs which are now
described. These two BSEs demonstrate that the Coulomb potential cannot mix
the $\Psi_{Atom}$ and $\Psi_{An}$ states.

These two different BSEs are derived from the two-body Green's functions for
the electron and positron propagators in the relative coordinates. Unlike the
TBDE, the BSE is relativistically covariant and treats the relative time
$t=t_{e}-t_{p}$ and the relative energy $\varepsilon=(e_{e}-e_{p})/2$ of the
two particles correctly as the fourth component of the coordinates
$\rho=(\boldsymbol{\rho},it)$ and conjugate momenta $\pi=(\boldsymbol{\pi
},i\varepsilon)$, respectively. One can define two propagators $K_{F}%
^{2}=K_{F}(e)\times K_{F}(p)$ and $K_{R}^{2}=K_{R}(e)\times K_{R}(p)$ for the
two-body Green's functions. For the Feynman propagator $K_{F}$, the positive
energy states are propagated forward in time, but the negative energy states
are propagated backward in time. For the retarded propagator $K_{R}$, both
positive and negative energy states are propagated forward in time. For the
two equations below, one needs to separate the atomic free -particle states
from the anomalous free -particle states as in (\ref{2.6}) and (\ref{2.7}).
For this purpose, the projection operators $\boldsymbol{\Lambda}_{\pm\pm}$ are
defined by $\boldsymbol{\Lambda}_{\pm\pm}(\boldsymbol{\pi})\Psi
(\boldsymbol{\pi})=\Psi_{\pm\pm}(\boldsymbol{\pi})$.

For atomic bound-states $\Psi_{Atom}$, it is necessary to use the two-body
Feynman propagator $K_{F}^{2}$ \cite{Salpeter1952} with the result that the
atomic states are only comprised of the free -particle wavefunctions
$\Psi_{_{++}}(\boldsymbol{\pi})$ and $\Psi_{--}(\boldsymbol{\pi})$ in
(\ref{2.6}). In the momentum representation, with a Coulomb potential
$\Phi_{C}$ , the BSE for $\Psi_{Atom}$ becomes, after some approximations
\cite{Patterson2019},
\begin{equation}
H_{0}[\Psi_{_{++}}(\boldsymbol{\pi})+\Psi_{--}(\boldsymbol{\pi})]-\frac{e^{2}%
}{2\pi^{2}}\boldsymbol{\Lambda}_{Atom}\int d^{3}k\frac{1}{\boldsymbol{k\cdot
k}}\Psi(\boldsymbol{k+\pi})=E[\Psi_{_{++}}(\boldsymbol{\pi})+\Psi
_{--}(\boldsymbol{\pi})], \label{3.1}%
\end{equation}
with $\boldsymbol{\Lambda}_{Atom}=\boldsymbol{\Lambda}_{++}(\boldsymbol{\pi
})-\boldsymbol{\Lambda}_{--}(\boldsymbol{\pi})$ and $\Psi_{+-}(\boldsymbol{\pi
})=\Psi_{-+}(\boldsymbol{\pi})=0$.

For the anomalous bound-states $\Psi_{DV}$, it is necessary to use the
two-body retarded propagator $K_{R}^{2}$ \cite{Patterson2019} with the result
that the $\Psi_{DV}$ states are only comprised of the anomalous states
$\Psi_{_{+-}}(\boldsymbol{\pi})$ and $\Psi_{-+}(\boldsymbol{\pi})$ where
$H_{0}\Psi_{+-}(\boldsymbol{\pi})=H_{0}\Psi_{-+}(\boldsymbol{\pi})=0$ in
(\ref{2.7}). Consequently, for a Coulomb potential, one finds the BSE for
$\Psi_{DV}$ \cite{Patterson2019} in (\ref{1.1}) and (\ref{1.3}) is
\begin{equation}
-\frac{e^{2}}{2\pi^{2}}\boldsymbol{\Lambda}_{DV}\int d^{3}k\frac
{1}{\boldsymbol{k\cdot k}}\Psi(\boldsymbol{k+\pi})=E[\Psi_{+-}(\boldsymbol{\pi
})+\Psi_{-+}(\boldsymbol{\pi})], \label{3.2}%
\end{equation}
with $\boldsymbol{\Lambda}_{DV}=\boldsymbol{\Lambda}_{+-}(\boldsymbol{\pi
})+\boldsymbol{\Lambda}_{-+}(\boldsymbol{\pi})$ and $\Psi_{_{++}%
}(\boldsymbol{\pi})=\Psi_{--}(\boldsymbol{\pi})=0$. Note that the BSE
(\ref{3.2}) for $\Psi_{DV}$ is simply the equivalent of diagonalizing
$\Phi_{C}$ in the free-particle bases $\Psi_{+-}(\boldsymbol{\pi})$ and
$\Psi_{-+}(\boldsymbol{\pi})$ of $\Psi_{An}$ corresponding to (\ref{1.3}).
This equation is not correct when the masses of the two particles are unequal
as in the case of Hydrogen where $H_{0}\Psi_{+-}(\boldsymbol{\pi})=-H_{0}%
\Psi_{-+}(\boldsymbol{\pi})\neq0$. Thus, it only applies to particles with
equal masses which obey (\ref{1.2}).

One should compare the BSEs (\ref{3.1}) and (\ref{3.2}) with the TBDE or Breit
equation in (\ref{2.11}). A consequence of the two BSEs (\ref{3.1}) and
(\ref{3.2}) is that the atomic states $\Psi_{Atom}$ and DV states $\Psi_{DV}$
cannot be mixed by the Coulomb potential in contrast to the TBDE equation
(\ref{2.11}). Summarizing, one must first eliminate the anomalous
free-particle states $\Psi_{+-}$ and $\Psi_{-+}$ in the TBDE (\ref{2.11}) in
order to calculate the atomic positronium bound-states. Conversely, one must
first eliminate the atomic free-particle states $\Psi_{++}$ and $\Psi_{--}$ in
the TBDE (\ref{2.11}) in order to calculate the\textbf{ }positronium DV states.

Mathematically, it must be emphasized that the BSE for $\Psi_{Atom}$
(\ref{3.1}) automatically specifies the $K_{F}$ propagator for the atomic
bound-states and the BSE for $\Psi_{DV}$ (\ref{3.2}) automatically specifies
the $K_{R}$ propagator for the anomalous bound-states. One has no choice for
the temporal boundary conditions as they are determined by these equations
unambiguously. The two different bases formed from atomic and DV states are
each mathematically complete sets spatially and therefore overlap: they are
not orthogonal spatially. However, they are orthogonal temporally because of
their different time behaviors as determined by their different propagators.
That is why atomic and DV states cannot be mixed by the instantaneous Coulomb potential.

Physically, the temporal boundary condition arising from the $K_{R}$
propagator is incorrect for free-particle states. As shown by Feynman
\cite{Feynman1949}, the propagator $K_{F}$ must be used for free electrons and
positrons in order to be consistent with the known behavior of these particles
undergoing scattering or annihilation. Otherwise, a state $\psi_{+}$ can be
scattered into a state $\psi_{-}$ and be lost. Furthermore, QED requires that
a state $\psi_{-}$ going backward in time becomes the antiparticle state
$\psi_{+}$ going forward in time. For this reason, anomalous states $\Psi
_{An}$, which are propagated by $K_{R}$, can only exist as bound-states
$\Psi_{DV}$ and not as free-particles. The anomalous bound-states $\Psi_{DV}$
cannot dissociate into free-particles due to the temporal boundary condition
imposed by $K_{R}$ on the solutions to (\ref{3.2}).

\section{Discrete Variable (DV) Representation}

In this section, it is shown explicitly that there are four different
anomalous bound-state solutions $\Psi_{DV}$ to (\ref{1.3}) or the equivalent
(\ref{3.2}), for any $J>0$ corresponding to four orthogonal spinor
combinations of $\mathbf{e}_{ij}\Omega_{S_{z}}^{S}.$ In this paper only
$J_{z}=0$ states are considered. These four bound-states have radially
quantized wavefunctions $\Psi_{DV}=G\delta(\rho-\rho_{i})$ for an
instantaneous Coulomb potential $\Phi(\rho)=\Phi_{C}(\rho)$ with energies
$E=\Phi(\rho_{i})=-e^{2}/\rho_{i}$. These quantized wavefunctions $\Psi_{DV}$
correspond to the DV representation as reviewed in \cite{Light2000}. According
to DV theory, DV states can be found either numerically, by diagonalizing the
potential $\Phi_{C}$ in a complete basis set, or analytically, by using the
completeness relation for the bases.

It will also be shown that, if one includes the appropriate instantaneous
magnetic potential $\Phi_{M}=e^{2}(\boldsymbol{\alpha}_{e}\cdot
\boldsymbol{\alpha}_{p})/\rho$, the DV states are also eigenstates of the
resulting effective potential is $\Phi=\Phi_{C}+\Phi_{M}$. With this
potential, the same four bound-states $\Psi_{DV}$ have energies $E_{i}^{1}=0$
for the $S=1$ doublet and $E_{i}^{0}=$ $2e^{2}/\rho_{i}$ for the $S=0$
doublet. In this section DV states are in the rest frame where
\textbf{$P^{\prime}$}$\mathbf{=0}$. The case for the DV states boosted to the
moving frame where $\mathbf{P}^{\prime}\mathbf{\neq0}$ will be considered in
section V.

In spherical coordinates $\mathbb{\rho}=(\rho,\theta,\phi)$, one can use the
momentum representation with $k=\pi$ for anomalous states, for a given total
angular momentum $J=L+S$ and projection $J_{z}=M+S_{z}$. The wavefunctions
$\Psi_{DV}$ are then comprised of states $j_{L}(k\rho)Y_{LM}(\theta
,\phi)e_{ij}\Omega_{S_{z}}^{S}$ which are products of spherical Bessel
functions $j_{L}(k\rho)$, spherical harmonics $Y_{LM}(\theta,\phi)$, and
spinors $e_{ij}\Omega_{S_{z}}^{S}$. The DV representation allows us to form
bound-states $\Psi_{DV}=G\delta(\rho-\rho_{i})$ using the completeness
condition on the spherical Bessel functions $j_{L}(k\rho)$ for the radial
wavefunctions. Using the boundary condition,%
\begin{equation}
j_{J}(k_{n}\rho_{0})=0, \label{4.9}%
\end{equation}
with the normalization%
\[
N_{Jn}=\sqrt{\frac{2}{\rho_{0}^{3}j_{J+1}^{2}(k_{n}\rho_{0})}}\simeq
\sqrt{\frac{2}{\rho_{0}}}k_{n}\text{ \ for }k_{n}>>0,
\]
one has the orthonormality condition%
\begin{equation}
N_{Jn}^{2}%
{\textstyle\int_{0}^{\rho_{0}}}
\rho^{2}j_{J}(k_{n}\rho)j_{J}(k_{m}\rho)d\rho=\delta_{nm}, \label{4.11}%
\end{equation}
and the completeness relation%
\begin{equation}
\sum_{n=1}^{\infty}N_{Jn}^{2}\rho^{2}j_{J}(k_{n}\rho)j_{J}(k_{n}\rho
_{i})=\delta(\rho-\rho_{i}). \label{4.12}%
\end{equation}

For a given $J$ and $k=\pi$, the DV states are denoted by $|\Psi^{S}%
,Jk\rangle$ for total spin $S=0$ and $S=1$. The anomalous states with
$\left\langle K\right\rangle =0$, which can form bound-states, are found to
be
\begin{subequations}
\label{4.1}%
\begin{align}
|\Psi_{A}^{0},Jk\rangle &  =\varphi_{0}(Jk)(\mathbf{e}_{11}-\mathbf{e}%
_{22})/\sqrt{2},\label{4.1a}\\
|\Psi_{S}^{0},Jk\rangle &  =\varphi_{0}(Jk)(\mathbf{e}_{12}-\mathbf{e}%
_{21})/\sqrt{2},\nonumber\\
|\Psi_{1S}^{1},Jk\rangle &  =\varphi_{1}(Jk)(\mathbf{e}_{11}+\mathbf{e}%
_{22})/\sqrt{2},\label{4.1b}\\
|\Psi_{2S}^{1},Jk\rangle &  =\varphi_{1}(Jk)(\mathbf{e}_{12}+\mathbf{e}%
_{21})/\sqrt{2},\nonumber
\end{align}
with the normalized functions,
\end{subequations}
\begin{subequations}
\label{4.2}%
\begin{align}
\varphi_{0}(Jk)  &  =N_{Jk}j_{J}(k\rho)[Y^{J}\Omega^{0}]_{0}^{J}%
,\label{4.2a}\\
\varphi_{1}(Jk)  &  =N_{Jk}j_{J}(k\rho)[Y^{J}\Omega^{1}]_{0}^{J}%
,\ \label{4.2b}%
\end{align}
and the Clebsch-Gordon $LS$ coupling,%
\end{subequations}
\[
\lbrack Y^{L}\Omega^{S}]_{0}^{J}=%
{\textstyle\sum\nolimits_{M}}
C_{M\ -M\ 0}^{L\ \ \ S\ \ J}Y_{M}^{L}\Omega_{-M}^{S}.
\]
For the pair of equations (\ref{4.1b}), which were not considered previously
in \cite{Patterson2019}, one must have $J>0$.

The subscripts $S$ and $A$ for these anomalous states indicate that these
wavefunctions are composed of symmetric states $(X=1)$ and antisymmetric
states $(X=-1)$, respectively, under particle exchange $\boldsymbol{X}%
\Psi=X\Psi$ where%
\begin{equation}
\Psi_{S}=\sqrt{%
\frac12
}(\Psi_{+-}+\Psi_{-+}),\ \ \Psi_{A}=\sqrt{%
\frac12
}(\Psi_{+-}-\Psi_{-+}),\ \ \label{4.4}%
\end{equation}
with $X$ being the negative of the charge conjugation $C$ or $X=-C$. The
exchange symmetry $X$ and inversion parity $P$ are given in Table I for these
four anomalous states. The labels $S$ and $A$ are for even $J$ only. For odd
$J$, the labels $S$ and $A$ will be reversed ($S\leftrightarrow A$).%
\begin{align*}
&  \text{Table I. Exchange }X=-C\text{ and Parity }P\\
&
\begin{tabular}
[c]{lllll}\hline\hline
& $|\Psi_{A}^{0},Jk\rangle$ & $\left\vert \Psi_{S}^{0},Jk\right\rangle $ &
$|\Psi_{1S}^{1},Jk\rangle$ & $|\Psi_{2S}^{1},Jk\rangle$\\\hline
$\ \ \ X$ & $(-1)^{J+1}$ & $(-1)^{J}$ & $(-1)^{J}$ & $(-1)^{J}$\\
$\ \ \ P$ & $(-1)^{J+1}$ & $(-1)^{J}$ & $(-1)^{J+1}$ & $(-1)^{J}%
$\\\hline\hline
\end{tabular}
\end{align*}
The labeling of these states agrees with that of Malenfant
\cite{Malenfant1988}. Note that, for a given $J$ and $k$ in Table I, the two
states $\Psi_{A}^{0}$ and $\Psi_{S}^{0}$ for the $S=0$ doublet have different
$X$ and $P$ and the two states $\Psi_{1S}^{1}$ and $\Psi_{2S}^{1}$ for the
$S=1$ doublet have different $P$. That is, neither $P$ nor $C$ is conserved
for a given $J$ and $k$. Also note that the $\Psi_{S}$ and the $\Psi_{A}$
states are their own antiparticles if one lets $\Psi_{+-}\leftrightarrow
\Psi_{-+}$.

It can be readily shown that the two pairs of states $|\Psi^{0},Jk\rangle$ and
$|\Psi^{1},Jk\rangle$ are each anomalous states $\Psi_{An}$ which obey
(\ref{1.2}). To show this, one uses the relations
\begin{align*}
(\boldsymbol{\sigma}_{e}\cdot\boldsymbol{\pi}\text{\textbf{\ }})\varphi
_{0}(Jk)  &  =-k\varphi_{\alpha}(Jk),\\
-(\boldsymbol{\sigma}_{p}\cdot\boldsymbol{\pi}\text{\textbf{\ }})\varphi
_{0}(Jk)  &  =-k\varphi_{\alpha}(Jk),\\
(\boldsymbol{\sigma}_{e}\cdot\boldsymbol{\pi}\text{\textbf{\ }})\varphi
_{1}(Jk)  &  =\ \ k\varphi_{\beta}(Jk),\\
-(\boldsymbol{\sigma}_{p}\cdot\boldsymbol{\pi}\text{\textbf{\ }})\varphi
_{1}(Jk)  &  =-k\varphi_{\beta}(Jk),
\end{align*}
where%
\begin{align*}
\varphi_{\alpha}(Jk)  &  =iN_{Jk}\{aj_{J+1}(k\rho)[Y^{J+1}\Omega^{1}]_{0}%
^{J}+bj_{J-1}(k\rho)[Y^{J-1}\Omega^{1}]_{0}^{J}\},\\
\varphi_{\beta}(Jk)  &  =iN_{Jk}\{-bj_{J+1}(k\rho)[Y^{J+1}\Omega^{1}]_{0}%
^{J}+aj_{J-1}(k\rho)[Y^{J-1}\Omega^{1}]_{0}^{J}\},
\end{align*}
with recoupling coefficients%
\[
a=\sqrt{\frac{J+1}{2J+1}},~b=\sqrt{\frac{J}{2J+1}}.
\]
These relations result in the equations for $K=(\boldsymbol{\alpha}%
_{e}-\boldsymbol{\alpha}_{p})\cdot\boldsymbol{\pi}$\textbf{\ }$\ $in
(\ref{2.4}) such that
\begin{subequations}
\label{4.5}%
\begin{align}
K\varphi_{0}(Jk)(\mathbf{e}_{11}-\mathbf{e}_{22})  &  =0,\label{4.5a}\\
K\varphi_{0}(Jk)(\mathbf{e}_{12}-\mathbf{e}_{21})  &  =0.\nonumber\\
K\varphi_{1}(Jk)(\mathbf{e}_{11}+\mathbf{e}_{22})  &  =0,\label{4.5b}\\
K\varphi_{1}(Jk)(\mathbf{e}_{12}+\mathbf{e}_{21})  &  =0.\nonumber
\end{align}
Note that these four equations depend only on the Pauli- and Dirac-spinor
exchange properties of $(\boldsymbol{\alpha}_{e}-\boldsymbol{\alpha}_{p})$ for
the four wavefunctions in (\ref{4.1}). For the high $k>>m$ needed to form the
DV bound-states at $fermi$ distances, one may ignore the negligible mass term
$\mathcal{M}$ in (\ref{2.4}). As a result of (\ref{4.5}) and (\ref{2.4}) with
\textbf{$P^{\prime}$}$\mathbf{=}\mathbf{0}$ and $\mathcal{M}=0$, the four
states $|\Psi,Jk\rangle$ above obey the equation for anomalous states in
(\ref{1.2}) or
\end{subequations}
\begin{equation}
\mathbf{H}_{0}|\Psi,Jk\rangle=K|\Psi,Jk\rangle=0. \label{4.3}%
\end{equation}
Combining (\ref{4.1}) and (\ref{4.2}), one now has the four anomalous states,
\begin{subequations}
\label{4.8}%
\begin{align}
|\Psi_{A}^{0},Jk\rangle &  =N_{Jk}\ j_{J}(k\rho)[Y^{J}\Omega^{0}]_{0}%
^{J}(\mathbf{e}_{11}-\mathbf{e}_{22})/\sqrt{2},\label{4.8a}\\
|\Psi_{S}^{0},Jk\rangle &  =N_{Jk\ }j_{J}(k\rho)[Y^{J}\Omega^{0}]_{0}%
^{J}(\mathbf{e}_{12}-\mathbf{e}_{21})/\sqrt{2},\nonumber\\
|\Psi_{1S}^{1},Jk\rangle &  =N_{Jk}\ j_{J}(k\rho)[Y^{J}\Omega^{1}]_{0}%
^{J}(\mathbf{e}_{11}+\mathbf{e}_{22})/\sqrt{2},~J>0,\label{4.8b}\\
|\Psi_{2S}^{1},Jk\rangle &  =N_{Jk}\ j_{J}(k\rho)[Y^{J}\Omega^{1}]_{0}%
^{J}(\mathbf{e}_{12}+\mathbf{e}_{21})/\sqrt{2},~J>0,\nonumber
\end{align}
which form a bases for two pairs of anomalous bound-states or DV states
labeled by the Pauli-spinors $S=0$ in (\ref{4.8a}) and $S=1$ in (\ref{4.8b}).
These pairs are symmetrized functions $\Psi_{u}$ $\ $for $S=0$ and $\Psi_{g}$
for $S=1$ as defined in (\ref{2.10}) and are only weakly coupled by the mass
operator $\mathcal{M}$.

It is now possible to form four different DV bound-states from the anomalous
states $|\Psi,Jk\rangle$ in (\ref{4.8}) using the DV representation for any
potential $\Phi(\rho)$. For each of these four states, one can diagonalize the
potential matrix%
\end{subequations}
\begin{equation}
\Phi_{nm}=\left\langle \Psi,Jk_{n}\left\vert \Phi(\rho)\right\vert \Psi
,Jk_{m}\right\rangle =N_{Jn}N_{Jm}%
{\textstyle\int_{0}^{\rho_{0}}}
\rho^{2}j_{J}(k_{n}\rho)\Phi(\rho)(j_{J}(k_{m}\rho)d\rho, \label{4.10}%
\end{equation}
to find the DV representation numerically. One may also find the DV states
analytically by using the truncated completeness relation for a finite basis
set. For a finite bases set with $N$ different $k_{n}$ in (\ref{4.9}), one
finds, analytically, the approximate radial delta functions,%
\begin{equation}
R_{i}(\rho)=D\sum_{n=1}^{N}N_{Jn}^{2}\rho^{2}j_{J}(k_{n}\rho)j_{J}(k_{n}%
\rho_{i})\simeq D\delta(\rho-\rho_{i}),\ \ \label{4.12a}%
\end{equation}
where the $\rho_{i}$ $(i=1,2,...,N)$ are at the $N$ zeros of the Bessel
function $j_{J}(k_{N+1}\rho)$. The normalization $D$ depends on $\rho_{i}$ and
the associated grid spacing $\Delta\rho_{i}$. Note for high $N$ one has
$\delta(0)\simeq1/\Delta\rho$ and%

\[
D^{2}\int\delta^{2}(\rho-\rho_{i})d\rho\simeq D^{2}\delta^{2}(0)\Delta
\rho\simeq D^{2}/\Delta\rho=1.
\]
One can then use the approximations,%
\begin{equation}
\Delta\rho\sim\rho_{0}/N,\ \ \rho_{i}\sim i\Delta\rho,\ \ \ D\sim\sqrt
{\Delta\rho}. \label{4.12b}%
\end{equation}
From (\ref{1.3}), the corresponding energies are%
\begin{equation}
E_{i}\simeq\Phi\left(  \rho_{i}\right)  . \label{4.21}%
\end{equation}
For $k>>m$ when $\rho_{0}\sim fermi$, one can ignore the mass coupling term
$\mathcal{M}$ in (\ref{2.4}) and (\ref{2.10a}) which justify the assumption
made previously to obtain (\ref{4.8}).

It will be convenient to define%
\begin{align}
\Omega_{+}^{1}  &  =\sqrt{%
\frac12
}(\Omega_{-1}^{1}e^{i\phi}+\Omega_{1}^{1}e^{-i\phi}),\label{4.13}\\
\Omega_{-}^{1}  &  =\sqrt{%
\frac12
}(\Omega_{-1}^{1}e^{i\phi}-\Omega_{1}^{1}e^{-i\phi}),\nonumber
\end{align}
and expand the $LS$ coupling in terms of the (normalized) Associated Legendre
Polynomials \cite{Abramowitz1972} $A_{0}^{J}P_{0}^{J}(\cos\theta)$ and
$A_{1}^{J}P_{1}^{J}(\cos\theta)$,%
\begin{align*}
\lbrack Y^{J}\Omega^{0}]_{0}^{J}  &  =A_{0}^{J}P_{0}^{J}(\cos\theta)\Omega
_{0}^{0}/\sqrt{2\pi},\\
\lbrack Y^{J}\Omega^{1}]_{0}^{J}  &  =A_{1}^{J}P_{1}^{J}(\cos\theta)\Omega
_{+}^{1}/\sqrt{2\pi}.
\end{align*}
For a given $J$, one obtains the DV bound-states $\Psi_{DV}$ from (\ref{4.8})
and (\ref{4.12a}) in terms of the Pauli- and Dirac-spinors,
\begin{subequations}
\label{4.40}%
\begin{align}
|\Psi_{A}^{0},J\rho_{i}\rangle &  =\Delta_{i}^{0}\Omega_{0}^{0}(\mathbf{e}%
_{11}-\mathbf{e}_{22})/\sqrt{2},\label{4.40a}\\
|\Psi_{S}^{0},J\rho_{i}\rangle &  =\Delta_{i}^{0}\Omega_{0}^{0}(\mathbf{e}%
_{12}-\mathbf{e}_{21})/\sqrt{2},\nonumber\\
|\Psi_{1S}^{1},J\rho_{i}\rangle &  =\Delta_{i}^{1}\Omega_{+}^{1}%
(\mathbf{e}_{11}+\mathbf{e}_{22})/\sqrt{2},\label{4.40b}\\
|\Psi_{2S}^{1},J\rho_{i}\rangle &  =\Delta_{i}^{1}\Omega_{+}^{1}%
(\mathbf{e}_{12}+\mathbf{e}_{21})/\sqrt{2},\nonumber
\end{align}
for $i=1,2,...,N$ in (\ref{4.12a}). The factors $\Delta_{i}$ for the
$(\rho,\theta)$ dependence, for a given $J$ and $\rho_{i},$ are%
\end{subequations}
\begin{align}
\Delta_{i}^{0}  &  \simeq D_{i}^{0}\delta(\rho-\rho_{i})A_{0}^{J}P_{0}%
^{J}(\cos\theta)/\sqrt{2\pi},\label{4.22}\\
\Delta_{i}^{1}  &  \simeq D_{i}^{1}\delta(\rho-\rho_{i})A_{1}^{J}P_{1}%
^{J}(\cos\theta)/\sqrt{2\pi},\nonumber
\end{align}
with the normalization $A_{0}^{J}$ and $A_{1}^{J}$ determined on the interval
$[-1,1]$ for $\cos\theta$, such that%
\begin{align*}
A_{0}^{J}  &  =\sqrt{\frac{(2J+1)}{2}},\\
A_{1}^{J}  &  =-\sqrt{\frac{(2J+1)}{2J(J+1)}}.
\end{align*}
The divisor $\sqrt{2\pi}$ is the normalization of the $\phi$ bases and the
normalization $D_{i}$ depend on the width $\Delta\rho_{i}$ of the delta
function $\delta(\rho-\rho_{i})$. For a finite basis set the delta functions
$\delta(\rho-\rho_{i})$ for $\Delta_{i}$ are only approximate, as indicated in
(\ref{4.22}).

\subsection{Coulomb Potential}

It is instructive to first use a Coulomb potential, $\Phi_{C}=-e^{2}/\rho$, in
order to examine the wavefunctions and energies of the four $\Psi_{DV}$ states
using DV theory. An example using DV theory for $J=10$ is shown in Fig. 1 for
the DV normalized wavefunctions $R_{i}^{2}(\rho)$ and in Fig. 2 for the DV
energies $E_{i}$ using a basis set with $N=40$ and $\rho_{0}=10^{-4}\ bohr$.
The wavefunctions and energies can be found analytically from (\ref{4.12a}),
(\ref{4.12b}), and (\ref{4.21}), using the known zeros $\rho_{i}$ of
$j_{10}(k_{41}\rho),$ or can be found numerically by the diagonalization of
the matrix $\Phi_{nm}$ from (\ref{4.10}). These figures show that the analytic
and numerical results are in agreement except for low $\rho_{i}$ where the
analytic approximations (\ref{4.12b}) fail for high $J$ and low $\rho_{i}$.%

\begin{figure}
[ptb]
\begin{center}
\ifcase\msipdfoutput
\includegraphics[
height=4.3465in,
width=4.8871in
]%
{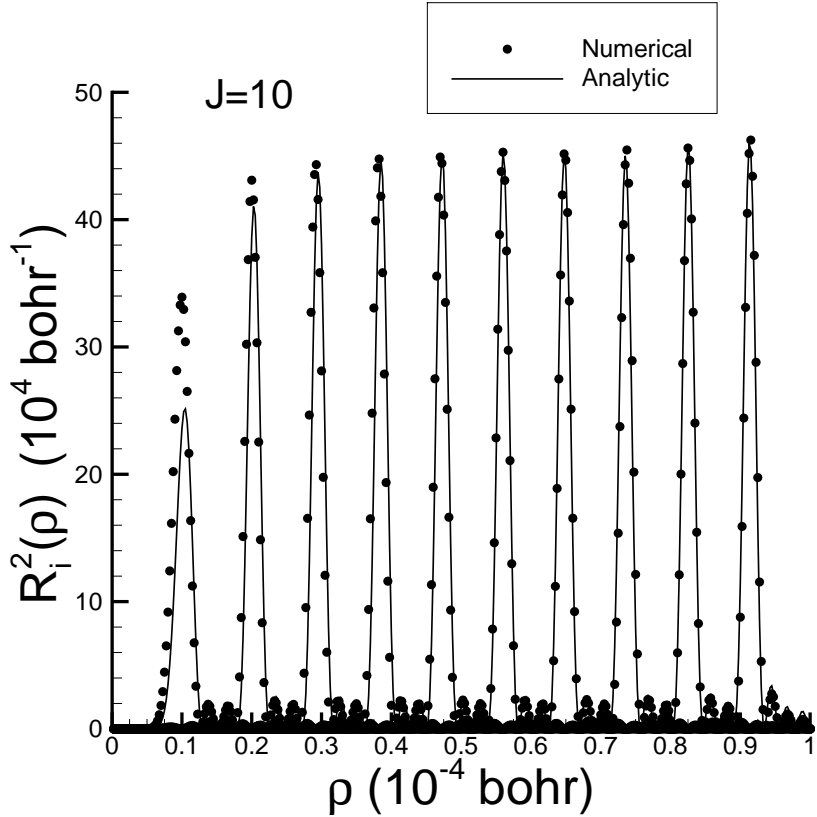}%
\else
\includegraphics[
height=4.3465in,
width=4.8871in
]%
{C:/Users/cwpat/Documents/Publications SWP/Neutrino 2022/arXiv/Arxiv 2/graphics/Fig1__1.pdf}%
\fi
\caption{The $R_{i}^{2}(\rho)$ DV wavefunctions for J=10 and N=40 with
$\rho_{0}=10^{-4}\ bohr.$ The numerical wavefunctions are calculated by
diagonalizing the potential matrix $\Phi_{nm}$ in (27) and the analytic
wavefunctions are found from (28) and (29). For clarity, only every fourth
wavefunction is shown.}%
\label{Fig.1}%
\end{center}
\end{figure}
%

\begin{figure}
[ptb]
\begin{center}
\ifcase\msipdfoutput
\includegraphics[
height=4.3465in,
width=4.8871in
]%
{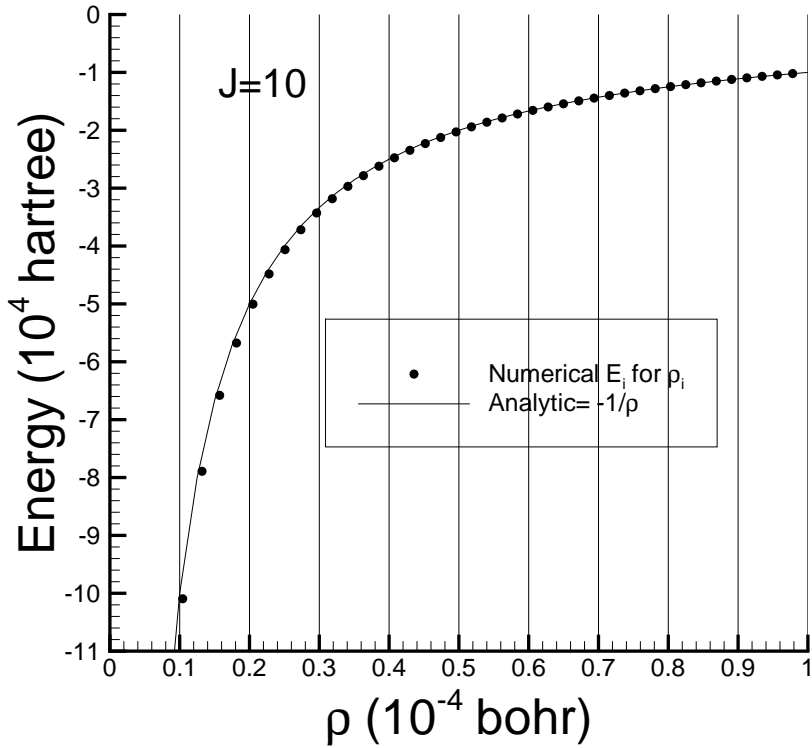}%
\else
\includegraphics[
height=4.3465in,
width=4.8871in
]%
{C:/Users/cwpat/Documents/Publications SWP/Neutrino 2022/arXiv/Arxiv 2/graphics/Fig2__2.pdf}%
\fi
\caption{The numerical energies $E_{i}$ are found by diagonalizing the
potential matrix $\Phi_{nm}$ in (27) with J=10 and N=40. The $\rho_{i}$ are
zeros of $j_{10}(k_{41}\rho)$.}%
\label{Fig.2}%
\end{center}
\end{figure}

\subsection{Magnetic Potential}

Using the Coulomb potential $\Phi_{C}(\rho)=-e^{2}/\rho$ in (\ref{1.1}) for
the DV states, one obtains the energies $E_{i}=-e^{2}/\rho$ for all four
$\Psi_{DV}$ in (\ref{4.40}). However, one still needs to include the
appropriate magnetic potential $\Phi_{M}$ so that the total Coulomb potential
$\Phi$ is then%
\begin{equation}
\Phi\mathbf{=}\Phi_{C}+\Phi_{M}. \label{5.0}%
\end{equation}
The appropriate magnetic potential will depend on the gauge used. While, in
theory, results should be gauge independent, in practice, different bases have
different QED convergence characteristics which make some gauge choices impractical.

For the atomic positronium states where $k<<m$, one finds that the expectation
value of the Dirac operators $\left\langle \boldsymbol{\alpha}_{e}%
\right\rangle =-\left\langle \boldsymbol{\alpha}_{p}\right\rangle $ is of
order $\alpha=e^{2}/\hbar c\sim1/137$. Here the weak-weak to strong-strong
component ratio $\Psi_{22}/\Psi_{11}$ is of order $\alpha^{4}.$ For the atomic
calculations, it is most convenient to use the Coulomb gauge with the Breit
magnetic potential $\Phi_{B}$ (derived from second order perturbation theory),
so that%
\begin{align*}
\Phi_{B}  &  =\frac{e^{2}}{2\rho}[\boldsymbol{\alpha}_{e}\cdot
\boldsymbol{\alpha}_{p}+\left(  \boldsymbol{\alpha}_{e}\cdot\widehat
{\boldsymbol{\rho}}\right)  \left(  \boldsymbol{\alpha}_{p}\cdot
\widehat{\boldsymbol{\rho}}\right)  ],\\
\Phi_{Atom}  &  =\Phi_{C}+\Phi_{B}=\Phi_{C}\{1-%
\frac12
[\boldsymbol{\alpha}_{e}\cdot\boldsymbol{\alpha}_{p}+\left(
\boldsymbol{\alpha}_{e}\cdot\widehat{\boldsymbol{\rho}}\right)  \left(
\boldsymbol{\alpha}_{p}\cdot\widehat{\boldsymbol{\rho}}\right)  ]\},
\end{align*}
which results in fine structure corrections to the Coulomb energies for
positronium. Using this magnetic potential and the Pauli approximation
restriction to a $\Psi_{++}$ bases, the fine structure energies for
positronium have been found analytically to order $m\alpha^{4}$ by Ferrell
\cite{Ferrell1951}, \cite{Bethe1957}. Also, Fulton and Martin
\cite{Fulton1954} have used the BSE for the $\Psi_{Atom}$ bases with various
two-body QED kernels in addition to the Coulomb potential to calculate the
energies of positronium to order $m\alpha^{5}$. In agreement with the BSE
(\ref{3.1}), they found it necessary to omit the $\Psi_{+-}$ and $\Psi_{-+}$
anomalous states. Also, unlike the TBDE, it is necessary to use the negative
of the Coulomb potential $-\Phi_{C}$ in $\Lambda_{Atom}$ for the negative
energy states $\Psi_{--}$.

For the DV positronium states in (\ref{4.40}), where $m<<k$, one finds that
the expectation value of the Dirac operators $\left\langle \boldsymbol{\alpha
}_{e}\right\rangle =\left\langle \boldsymbol{\alpha}_{p}\right\rangle
\rightarrow1$ when $\left\langle \mathcal{M}\right\rangle \rightarrow0$ in
(\ref{2.4}) and the components $\Psi_{11}$ and $\Psi_{22}$ are equal in
magnitude as shown in (\ref{4.40}).$~$Because of these extremely relativistic
states, QED covariant perturbation theory is not convergent and
diagonalization of the potential is necessary. Indeed, one finds that the
expectation values of the magnetic and electromagnetic potentials become
comparable in magnitude as in the classical case for highly relativistic
particles. Accordingly, one must use the DV potential $\Phi_{DV}$ with the
instantaneous magnetic potential of Gaunt $\Phi_{G}$ \cite{Alstine1997}, so
that%
\begin{align}
\Phi_{G}  &  =\frac{e^{2}}{\rho}\boldsymbol{\alpha}_{e}\cdot\boldsymbol{\alpha
}_{p}\equiv\frac{e^{2}}{\rho}(\alpha_{ex}\alpha_{px}+\alpha_{ey}\alpha
_{py}+\alpha_{ez}\alpha_{pz}),\label{5.1}\\
\Phi_{DV}  &  =\Phi_{C}+\Phi_{G}=\Phi_{C}\alpha_{0}^{2},\nonumber\\
\alpha_{0}^{2}  &  =(1-\boldsymbol{\alpha}_{e}\cdot\boldsymbol{\alpha}%
_{p}).\nonumber
\end{align}
This total potential has also been derived by Barut and Komy \cite{Barut1985a}
using the action principal. The operator $\boldsymbol{\alpha}_{e}%
\cdot\boldsymbol{\alpha}_{p}$ only acts on the Pauli- and Dirac-spinors of
$\Psi_{DV}$ in (\ref{4.40}) and it is shown below that the DV states are
eigenstates of $\alpha_{0}^{2}$ in (\ref{5.1}) as well as the potential
$\Phi_{C}$. Barut and Komy have shown that this instantaneous effective
potential is appropriate for the retarded propagator $K_{R}$ which is the
propagator used for the DV states in (\ref{3.2}). As described in the
Appendix, the DV potential $\Phi_{DV}=\Phi_{C}\alpha_{0}^{2}$ is related to
the Lorentz potential
\begin{equation}
\Phi_{L}=\Phi_{C}\gamma_{0}^{2}, \label{5.8}%
\end{equation}
where $\gamma_{0}^{2}=$ $\gamma_{eu}\gamma_{pu}$ is the scalar product of
$\gamma_{eu}$ and $\gamma_{pu}$. The correct potential for the BSE equation in
(\ref{3.2}), which uses the $K_{R}$ propagator, should be multiplied on the
left by the factor $\alpha_{0}^{2}$.

One can evaluate the Pauli-spinor operator $\boldsymbol{\sigma}_{e}%
\cdot\boldsymbol{\sigma}_{p}$ in the magnetic potential for states in
(\ref{4.40}) using%
\begin{align}
(\boldsymbol{\sigma}_{e}\cdot\boldsymbol{\sigma}_{p})\Omega_{0}^{0}  &
=-3\Omega_{0}^{0},\label{5.3}\\
(\boldsymbol{\sigma}_{e}\cdot\boldsymbol{\sigma}_{p})\Omega_{S_{z}}^{1}  &
=\Omega_{S_{z}}^{1},\nonumber
\end{align}
for $S_{z}=-1,0,1$. Evaluating the operator $\boldsymbol{\alpha}_{e}%
\cdot\boldsymbol{\alpha}_{p}$ for Dirac- and Pauli-spinor states gives%
\begin{align}
(\boldsymbol{\alpha}_{e}\cdot\boldsymbol{\alpha}_{p})(\mathbf{e}_{11}%
\pm\mathbf{e}_{22})\Omega_{0}^{0}  &  =\mp3(\mathbf{e}_{11}\pm\mathbf{e}%
_{22})\Omega_{0}^{0},\label{5.4}\\
(\boldsymbol{\alpha}_{e}\cdot\boldsymbol{\alpha}_{p})(\mathbf{e}_{12}%
\pm\mathbf{e}_{21})\Omega_{0}^{0}  &  =\mp3(\mathbf{e}_{12}\pm\mathbf{e}%
_{21})\Omega_{0}^{0},\nonumber\\
(\boldsymbol{\alpha}_{e}\cdot\boldsymbol{\alpha}_{p})(\mathbf{e}_{11}%
\pm\mathbf{e}_{22})\Omega_{S_{z}}^{1}  &  =\pm(\mathbf{e}_{11}\pm
\mathbf{e}_{22})\Omega_{S_{z}}^{1},\nonumber\\
(\boldsymbol{\alpha}_{e}\cdot\boldsymbol{\alpha}_{p})(\mathbf{e}_{12}%
\pm\mathbf{e}_{21})\Omega_{S_{z}}^{1}  &  =\pm(\mathbf{e}_{11}\pm
\mathbf{e}_{22})\Omega_{S_{z}}^{1},\nonumber
\end{align}
so that these spinor states are all eigenfunctions of $\boldsymbol{\alpha}%
_{e}\cdot\boldsymbol{\alpha}_{p}$. For the DV states in (\ref{4.40}), one then
finds%
\begin{align}
\Phi_{DV}|\Psi^{0},J\rho_{i}\rangle &  =E_{i}^{0}|\Psi^{0},J\rho_{i}%
\rangle,\label{5.2}\\
E_{i}^{0}  &  =M_{i}^{0}=2e^{2}/\rho_{i},\nonumber\\
\Phi_{DV}|\Psi^{1},J\rho_{i}\rangle &  =E_{i}^{1}|\Psi^{1},J\rho_{i}%
\rangle,\nonumber\\
E_{i}^{1}  &  =M_{i}^{1}=0.\nonumber
\end{align}
These DV eigenstates of $\Phi_{DV}$ no longer have negative energies and are,
therefore, physically allowed.

One can divide the magnetic Gaunt potential into its longitudinal and
transverse components,%
\begin{equation}
\Phi_{G}=\Phi_{GL}+\Phi_{GT}, \label{5.0a}%
\end{equation}
where%
\begin{equation}
\Phi_{GL}=\frac{e^{2}}{\rho}(\alpha_{ez}\alpha_{pz}),\ \ \ \Phi_{GT}%
=\frac{e^{2}}{\rho}(\alpha_{ex}\alpha_{px}+\alpha_{ey}\alpha_{py}).
\label{5.5}%
\end{equation}
Interestingly, one has for all DV states in (\ref{4.40}),%
\begin{equation}
\alpha_{ez}\alpha_{pz}|\Psi,J\rho_{i}\rangle=|\Psi,J\rho_{i}\rangle,
\label{5.6}%
\end{equation}
so that%
\begin{align}
(\Phi_{C}+\Phi_{GL})|\Psi,J\rho_{i}\rangle &  =0,\label{5.7}\\
\Phi_{DV}|\Psi,J\rho_{i}\rangle &  =\Phi_{GT}|\Psi,J\rho_{i}\rangle.\nonumber
\end{align}
This means that, for the DV bound-states, only the transverse magnetic
potential $\Phi_{GT}$ of Gaunt determines the energies in (\ref{5.2}). On the
other hand, for the atomic states, it is the Coulomb potential which is dominant.

Summarizing, the DV states $|\Psi^{0},J\rho_{i}\rangle$ for $\Omega_{0}^{0}$
form `heavy' doublets with energy $E_{i}^{0}=2e^{2}/\rho_{i}$ and the DV
states $|\Psi^{1},J\rho_{i}\rangle$ for $\Omega_{+}^{1}$ form `light' doublets
with $E_{i}^{1}=0$. The four degenerate states in (\ref{4.40}) with Coulomb
energies $E_{i}=-e^{2}/\rho_{i}$ are now split by the magnetic potential into
two different doublets.

\section{Classical Correspondence}

For clarity, the constant $c$ is now shown in this section. The different
behaviors between the atomic and anomalous states all depend on the fact that,
for anomalous states, either the electron or positron is in a negative energy
state $\psi_{-}(e)$ or $\psi_{-}(p)$, respectively. It is possible to
understand the important properties of the DV states by using both quantum and
classical principals. In particular, it is now shown that the four anomalous
bound-states $\Psi_{DV}$ are both stable and dark, as was the case for the
$J=0$ bound-states in \cite{Patterson2019}.

For a free-particle with energy $E=\pm\left\vert E\right\vert $ and momentum
$\mathbf{p}$, one obtains the important result \cite{Sakurai1967} for the
expectation value of the Dirac matrices $\alpha_{k}$,
\begin{equation}
\left\langle \alpha_{k}\right\rangle =p_{k}c/E=\pm p_{k}c/\left\vert
E\right\vert =v_{k}/c. \label{9.1}%
\end{equation}
In other words, the velocity $\mathbf{v}$ is in the direction opposite to the
momentum $\mathbf{p}$ for negative energy states of free-particles. In the
\textbf{$P^{\prime}$}$=\mathbf{0}$ frame, where $\mathbf{p}_{e}=-\mathbf{p}%
_{p}$, one finds that $\mathbf{v}_{e}=\mathbf{v}_{p}$ because either the
electron or the positron is in a negative energy state. The difference in
velocity components is then
\begin{equation}
\left\langle (\boldsymbol{\alpha}_{e}-\boldsymbol{\alpha}_{p})_{k}%
\right\rangle =0\mathbf{.} \label{9.3}%
\end{equation}
One can now understand why kinetic energy $\left\langle K\right\rangle $ is
zero in (\ref{4.5}) despite the fact that the relativistic momentum $\pi$ is
large, $\pi=k>>mc$. One has%
\begin{align*}
\left\langle K\right\rangle  &  =c\left\langle (\boldsymbol{\alpha}%
_{e}-\boldsymbol{\alpha}_{p})\cdot\mathbf{\pi}\right\rangle ,\\
&  =c\left\langle (\boldsymbol{\alpha}_{e}-\boldsymbol{\alpha}_{p}%
)_{k}\right\rangle \left\langle \mathbf{\pi}_{k}\right\rangle =0.
\end{align*}
Classically, (\ref{9.1}) corresponds to $m=-\left\vert m\right\vert $ when
$E=-\left\vert E\right\vert $ and for $\mathbf{p}_{e}=-\mathbf{p}_{p}$ one
also finds that $\mathbf{v}_{e}=\mathbf{v}_{p}$ for both the anomalous states
$\Psi_{+-}$ or $\Psi_{-+}$.

With this in mind, one can also understand why the DV states are delta
functions with $\Psi_{DV}=G\delta(\rho-\rho_{i})$. From the Heisenberg
uncertainty principal, these delta functions correspond to very high momentum
states where $\pi>>mc$. Because the relative velocity is zero, $\mathbf{v}%
_{e}-\mathbf{v}_{p}=0$, the classical particles will remain at the same
arbitrary distance $\rho=\rho_{i}$. Also, the electron and positron particles
can never annihilate because they can never collide. Finally, two particles
moving at the same velocity can neither radiate nor absorb light. This later
property is also clear from quantum considerations, given that the expectation
value of the Dirac transition operator $\mathbf{W}_{\mu}$ is%
\begin{equation}
\left\langle \mathbf{W}_{\mu}\right\rangle =-e\left\langle (\boldsymbol{\alpha
}_{e}-\boldsymbol{\alpha}_{p})\cdot\mathbf{1}_{\mu}\right\rangle =0,
\end{equation}
where $\mu=z$ for longitudinal and $\mu=x,y$ for transverse radiation.

It is also apparent that an electron and positron moving with the same
velocity can keep the same time so that one may let $t_{e}=t_{p}=t$. For this
reason, there will be no difficulties with the DV solutions of the BSE in
(\ref{3.2}) that arise from the particles keeping different times. The total
time $T$ is the fourth component of $\mathbf{R}$ in (\ref{2.2}) and is
conjugate to the total energy $E$. One sees that the total time $T$ and the
relative time $t$ are the same,%
\begin{equation}
T=\frac{t_{e}+t_{p}}{2}=t.
\end{equation}
In many respects, the solutions to the BSE (\ref{3.2}) for anomalous
bound-states $\Psi_{DV}$ are simpler than the solutions to the BSE (\ref{3.1})
for the atomic states $\Psi_{Atom}$.

Finally, one can readily determine the classical instantaneous electromagnetic
interaction between electron and positron point particles with angular
momentum $J\neq0$. \ One finds the magnetic force $F_{M}$ is%

\begin{equation}
F_{M}=-\frac{e^{2}}{\rho^{2}c^{2}}[\mathbf{v}_{e}\times(\mathbf{v}_{p}%
\times\widehat{\boldsymbol{\rho}})]
\end{equation}
for velocities $v\ll c$. In the center of mass frame with $\mathbf{R}=0$, let
the classical velocities be perpendicular to $\mathbf{r}_{e}=\boldsymbol{\rho
}/2$ and $\mathbf{r}_{p}=-\boldsymbol{\rho}/2$ for $J\neq0$\ so that%

\[
\lbrack\mathbf{v}_{e}\times(\mathbf{v}_{p}\times\widehat{\boldsymbol{\rho}%
})]=(\mathbf{v}_{e}\cdot\widehat{\boldsymbol{\rho}})\mathbf{v}_{p}%
-\mathbf{(v}_{e}\mathbf{\cdot v}_{p}\mathbf{)}\widehat{\boldsymbol{\rho}%
}=-\mathbf{(v}_{e}\mathbf{\cdot v}_{p}\mathbf{)}\widehat{\boldsymbol{\rho}},
\]
and%
\begin{equation}
F_{M}=\frac{e^{2}}{\rho^{2}c^{2}}\mathbf{(v}_{e}\mathbf{\cdot v}_{p}%
\mathbf{)}\widehat{\boldsymbol{\rho}}.
\end{equation}
One finds that the force $F_{M}$ is repulsive in the $\widehat
{\boldsymbol{\rho}}$ direction for the electron and positron moving in the
same direction as required. The electromagnetic potential for $J\neq0$ is then%
\begin{equation}
\Phi=\Phi_{C}+\Phi_{M}=-\frac{e^{2}}{\rho}(1-\frac{\mathbf{v}_{e}\mathbf{\cdot
v}_{p}}{c^{2}}\mathbf{).} \label{9.4}%
\end{equation}

For relativistic velocities, the retarded Lienard-Wiechert potential preserves
the ratio $\Phi_{C}/\Phi_{M}=-\mathbf{v}_{e}\mathbf{\cdot v}_{p}/c^{2}$.
Letting $\mathbf{v}_{e}=\mathbf{v}_{p}=\mathbf{c}$ where $\Phi_{C}/\Phi
_{M}=-1$, the classical and quantum potential is then $\Phi=0$ as in
(\ref{5.2}) for the $S=1$ case. Note that the quantum potential (\ref{5.1})
can be derived from (\ref{9.4}) by replacing $\mathbf{v}_{e}\mathbf{\cdot
v}_{p}/c^{2}$ with $\boldsymbol{\alpha}_{e}\cdot\boldsymbol{\alpha}_{p}$. This
quantum potential $\Phi_{DV}=\Phi_{C}\alpha_{0}^{2}$ is valid for all
velocities and is in agreement with Barut and Komy \cite{Barut1985a} using the
retarded $K_{R}$ propagator. It's transformation to a moving frame is shown in
the Appendix.

\section{Lorentz Boosts, Dynamical Symmetry Breaking, and Majorana Fermions}

Consider a new frame in which the DV states are moving with velocity
$V^{\prime}$ in the $Z$ direction so that the DV states have total momentum
$P^{\prime}=P_{Z}^{\prime}$ relative to the rest frame. One can also, without
loss of generality, define the $z$ direction to be in the direction of this
momentum so that the components of spins $s_{z}$ will be defined in the
$\widehat{\mathbf{z}}=\widehat{\mathbf{Z}}$ direction.

One can transform the DV states $\Psi_{i}$ and potential $\Phi_{DV}=\Phi
_{C}\alpha_{0}^{2}$ (\ref{5.1}) to the moving frame using both the two-body
Lorentz boost $L^{2}$ of the Dirac spinors and the Lorentz contraction of the
$z$ coordinate. In the Appendix, some useful identities (\ref{12.5}) and
(\ref{12.8}) for the conjugate expectations of $\alpha_{0}^{2}$, $\Phi
_{C}\alpha_{0}^{2}$, and $\Phi_{C}^{\prime}\alpha_{0}^{2}$ are%
\begin{align}
\left\langle \Psi_{i}\left\vert \alpha_{0}^{2}\right\vert \Psi_{i}%
\right\rangle  &  =\left\langle \Psi_{i}^{\prime}\left\vert \alpha_{0}%
^{2}\right\vert \Psi_{i}^{\prime}\right\rangle ,\label{10.1a}\\
\langle\Psi_{i}|\Phi_{C}\alpha_{0}^{2}|\Psi_{i}\rangle &  =M_{i},\nonumber\\
\langle\Psi_{i}^{\prime}|\Phi_{C}^{\prime}\alpha_{0}^{2}|\Psi_{i}^{\prime
}\rangle &  =M_{i}^{\prime},\nonumber
\end{align}
where $M_{i}$ is the binding energy of $\Psi_{i}$ in the rest frame and
$M_{i}^{\prime}$ is the binding energy of $\Psi_{i}^{\prime}$ in the moving
frame. Only if this binding energy remains constant, such that $M_{i}^{\prime
}=M_{i}$, can one demonstrate that the DV states transform like
single-particle fermions. As shown below, the mass $M_{i}^{\prime}$ is a
constant in the moving frame for all $\Psi_{i}^{\prime1}$ DV states in
(\ref{4.40}), but only for a special case of the DV states $\Psi_{i}^{\prime
0}$ which corresponds to the lowest energy state.

The Lorentz boost has the properties%
\begin{align}
\Psi^{\prime}  &  =L^{2}\Psi,\ \Phi_{DV}^{\prime}=L^{2}\Phi_{DV}L^{-2}%
,\qquad\label{10.1}\\
L^{-2}  &  =\gamma_{4}^{2}L^{2}\gamma_{4}^{2},\ \ \ \gamma_{4}^{2}=\gamma
_{e4}\gamma_{p4},\ \nonumber
\end{align}
where $L^{-2}$ is the inverse transform of $L^{2}$. These transformations
involve some difficulties arising from dynamical symmetry breaking and the
fact that $L^{2}$ is not unitary, which are now addressed.

As shown by (\ref{4.40}) and (\ref{4.22}), the DV wavefunctions $\Psi_{i}^{0}$
and $\Psi_{i}^{1}$ for a given $J$ can be separated into two factors. One
factor consist only of the Dirac spinors $\Omega_{S_{z}}^{S}\mathbf{e}_{ij}$
and the other factor consists only of the coordinate functions $\Delta
_{i}(\rho,\theta)$. The Lorentz boost operates on the Dirac spinors
$\Omega_{S_{z}}^{S}\mathbf{e}_{ij}$, whereas the Lorentz contraction operates
on the functions $\Delta_{i}(\rho,\theta)$. Similarly, for the potential
$\Phi_{DV}$, the Lorentz boost operates on the factor $\alpha_{0}^{2}$ and the
Lorentz contraction operates on the factor $\Phi_{C}=-e^{2}/\rho$. The Lorentz
contraction of $\rho$ in $\Delta_{i}(\rho,\theta)$ and $\Phi_{C}$ will be
considered first.

Because of the delta function factor $\delta(\rho-\rho_{i})$ (\ref{4.22}), the
\ $\Delta_{i}(\rho,\theta)$ in the rest frame are eigenstates of $1/\rho$ in
$\Phi_{C}$ such that%
\begin{equation}
\frac{1}{\rho}\Delta_{i}(\rho,\theta)=\frac{1}{\rho_{i}}\Delta_{i}(\rho
,\theta). \label{10.3a}%
\end{equation}

However, as a result of the Lorentz contraction of $z_{i}$ in the potential,
$z_{i}^{\prime}=z_{i}/\gamma$, the separations $\rho_{i}^{\prime}$ are no
longer spherically symmetric. Because of this dynamical symmetry breaking, the
potential $\Phi_{C}$ becomes cylindrically symmetric and $\rho_{i}^{\prime}$
depends on $\theta_{i}^{\prime}$. This means that $\Delta_{i}^{\prime}%
(\rho^{\prime},\theta^{\prime})$ is no longer an eigenstate of the
$1/\rho^{\prime}$. This problem can be remedied by using delta functions
$\delta(\cos\theta-\cos\theta_{i})$ in $\Delta_{i}$ formed from the degenerate
$P_{0}^{J}(\cos\theta)$ and $P_{1}^{J}(\cos\theta)$ in (\ref{4.22}) without
any effect on the energies $E_{i}$. Letting $\zeta=\cos\theta$, the delta
functions for the DV states directed at angle $\theta=\theta_{i}$ are given by%

\begin{align}
\left\{
\begin{array}
[c]{c}%
\lbrack\delta(\zeta-\zeta_{i})+\delta(\zeta+\zeta_{i})\ \text{ }J=even\\
\lbrack\delta(\zeta-\zeta_{i})-\delta(\zeta+\zeta_{i})\ \text{ }J=odd
\end{array}
\right\}   &  =%
{\textstyle\sum\limits_{J}}
(2J+1)P_{0}^{J}(\zeta_{i})P_{0}^{J}(\zeta),\label{10.4}\\
\left\{
\begin{array}
[c]{c}%
\lbrack\delta(\zeta-\zeta_{i})+\delta(\zeta+\zeta_{i})\ \text{ }J=odd\\
\lbrack\delta(\zeta-\zeta_{i})-\delta(\zeta+\zeta_{i})\ \text{ }J=even
\end{array}
\right\}   &  =%
{\textstyle\sum\limits_{J}}
\frac{2J+1}{J(J+1)}P_{1}^{J}(\zeta_{i})P_{1}^{J}(\zeta).\nonumber
\end{align}
One must keep in mind that the peaks of the delta functions $\zeta_{i}$ are
not the same for the symmetric and antisymmetric delta functions above because
the peaks of $P_{0}^{J}(\zeta_{i})$ are interleaved for $J=even$ and $J=odd$
and similarly for $P_{1}^{J}(\zeta_{i})$. For example, the symmetric delta
functions can have a peak at $\zeta_{i}=0$ whereas the antisymmetric delta
functions cannot. This means that when $\zeta_{i}=0$, the $\Psi_{i}^{0}$ DV
state has $J=even$ and the $\Psi_{i}^{1}$ DV state has $J=odd$. In
(\ref{4.22}), one now has, approximately,
\begin{align}
\Delta_{i}^{0}  &  \simeq D_{i}^{0}\delta(\rho-\rho_{i})[\delta(\zeta
-\zeta_{i})\pm\delta(\zeta+\zeta_{i})]/\sqrt{4\pi},\label{4.22a}\\
\Delta_{i}^{1}  &  \simeq D_{i}^{1}\delta(\rho-\rho_{i})[\delta(\zeta
-\zeta_{i})\pm\delta(\zeta+\zeta_{i})]/\sqrt{4\pi,}\nonumber
\end{align}
from (\ref{10.4}). These new functions $\Delta_{i}$ are localized at the
coordinates $(\rho_{i},\theta_{i})$ and remain eigenstates of $1/\rho$
(\ref{10.3a}) because $\rho$ is independent of $\theta$.

For a given velocity $V^{\prime}$ with the Lorentz factor $\gamma$, one has
the following relations for the Lorentz contraction of $z$, using the
cylindrical coordinates $(r_{i},z_{i})$,%
\begin{align*}
\rho_{i}^{\prime}\cos\theta_{i}^{\prime}  &  =z_{i}^{\prime}=z_{i}/\gamma
=\rho_{i}\cos\theta_{i}/\gamma,\\
\rho_{i}^{\prime}\sin\theta_{i}^{\prime}  &  =r_{i}^{\prime}=r_{i}=\rho
_{i}\sin\theta_{i}.
\end{align*}
Solving these equations, one finds
\begin{align}
\rho_{i}^{\prime}  &  =(\rho_{i}/\gamma)\sqrt{\gamma^{2}-\cos^{2}(\theta
_{i})(\gamma^{2}-1)},\label{10.2}\\
\cos\theta_{i}^{\prime}  &  =\rho_{i}\cos\theta_{i}/(\rho_{i}^{\prime}%
\gamma).\nonumber
\end{align}
The new factors $\Delta_{i}^{\prime}$ for $\Psi_{i}^{\prime}$, resulting from
the Lorentz contraction, are then simply%
\begin{align}
\Delta_{i}^{0\prime}  &  \simeq D_{i}^{0\prime}\delta(\rho^{\prime}-\rho
_{i}^{\prime})[\delta(\zeta^{\prime}-\zeta_{i}^{\prime})\pm\delta
(\zeta^{\prime}+\zeta_{i}^{\prime})]/\sqrt{4\pi},\label{4.22b}\\
\Delta_{i}^{1\prime}  &  \simeq D_{i}^{1\prime}\delta(\rho^{\prime}-\rho
_{i}^{\prime})[\delta(\zeta^{\prime}-\zeta_{i}^{\prime})\pm\delta
(\zeta^{\prime}+\zeta_{i}^{\prime})]/\sqrt{4\pi},\nonumber
\end{align}
for DV states localized at coordinates $(\rho_{i}^{\prime},\theta_{i}^{\prime
})$ given by (\ref{10.2}) where $\zeta_{i}^{\prime}=\cos\theta_{i}^{\prime}$.
With this simple procedure, the factors $\Delta_{i}^{\prime}$ for $\Psi
_{i}^{\prime}$ are now eigenstates of $1/\rho^{\prime}$
\begin{equation}
\frac{1}{\rho^{\prime}}\Delta_{i}^{\prime}(\rho^{\prime},\theta^{\prime
})=\frac{1}{\rho_{i}^{\prime}}\Delta_{i}^{\prime}(\rho^{\prime},\theta
^{\prime}). \label{10.3b}%
\end{equation}

Now consider the Lorentz boost of $K,$ $\Phi_{DV}$, and the DV states
$\Psi_{i}$. In the Appendix it is shown, using adjoint spinors, that%

\[
\langle\Psi_{i}^{0\prime}|K^{\prime}|\Psi_{i}^{0\prime}\rangle=0,
\]
so the transformed wavefunctions are also DV states. For the potential, one
can also verify from (\ref{5.2}), (\ref{10.1a}), and (\ref{10.3b}) that%

\begin{align}
\langle\Psi_{i}^{0\prime}|\Phi_{C}^{\prime}\alpha_{0}^{2}|\Psi_{i}^{0\prime
}\rangle &  =M_{i}^{0\prime}=2e^{2}/\rho_{i}^{\prime},\label{5.2b}\\
\langle\Psi_{i}^{1\prime}|\Phi_{C}^{\prime}\alpha_{0}^{2}|\Psi_{i}^{1\prime
}\rangle &  =M_{i}^{1\prime}=0.\nonumber
\end{align}
For the $\Psi_{i}^{1\prime}$ DV states, one has $M_{i}^{1\prime}=M_{i}^{1}=0$
so that $\beta^{\prime}=1$ and the velocity $V^{\prime}$ will be the speed of
light. For these light DV states, one has $\gamma^{\prime}\rightarrow\infty$
and $z_{i}^{\prime}=0$ such that $\rho_{i}^{\prime}=\rho_{i}=r_{i}$ and they
form a disk perpendicular to the direction of motion $\mathbf{\hat{z}%
=}\widehat{\mathbf{Z}}$. In this case the potential expectation value
$\left\langle \Phi_{DV}\right\rangle $ is a Lorentz invariant as required for
these DV states to transform like a single-particle fermion.

For the $\Psi_{i}^{0\prime}$ DV states, one has $M_{i}^{0\prime}=2e^{2}%
/\rho_{i}^{\prime}\neq M_{i}^{0}$ when $z_{i}\neq0$ and the masses are not
Lorentz invariant. In fact one finds from (\ref{10.2}), for a given $\rho_{i}%
$, that $M_{i}^{0\prime}$ is a maximum when $\theta_{i}=0$ or $\theta_{i}=\pi$
where $\rho_{i}^{\prime}=\rho_{i}/\gamma$ and $M_{i}^{0\prime}=\gamma
M_{i}^{0}$. One also finds from (\ref{10.2}), for a given $\rho_{i}$, that
$M_{i}^{0\prime}$ is a minimum when $\theta_{i}=\pi/2$ $(z_{i}=0)$ where
$\rho_{i}^{\prime}=\rho_{i}$ and $M_{i}^{0\prime}=M_{i}^{0}$. In general one
finds that
\[
\gamma M_{i}^{0}\geq M_{i}^{0\prime}\geq M_{i}^{0}%
\]
for a given $\rho_{i}$. This is a consequence of the fact that, for a given
$\rho_{i},$ the Lorentz contraction will bring the electron and positron
closer together for any angle $\theta_{i}$ unless $\theta_{i}=\pi/2$. Because
the potential is repulsive, this will then raise the binding energy and mass
$M_{i}^{0\prime}$ unless $\theta_{i}=\pi/2$. For the special case where
$\theta_{i}=\pi/2$ $(z_{i}=0),$ the DV states $\Psi_{i}^{0\prime}$ form a disk
perpendicular to the direction of motion $\mathbf{\hat{z}=}\widehat
{\mathbf{Z}}$ such like the DV states $\Psi_{i}^{1\prime}$. One would expect
that this particle would reside in the lowest energy DV state of $\Psi
_{i}^{0\prime}$\ with $\theta_{i}=\pi/2$, where the mass is invariant. The
fact that such DV doublets $\Psi_{i}$ with $M_{i}^{0\prime}=M_{i}^{0}$
transform like single-particle fermions will now be shown.

Having transformed the coordinate factors $\Delta_{i}(\rho,\theta)$ for the DV
states $\Psi_{i}^{0}$ and $\Psi_{i}^{1}$ for $J=(even,odd)$ as shown in
(\ref{4.22b}), it remains to transform their Dirac spinors $\Omega_{S_{z}}%
^{S}\mathbf{e}_{ij}$ to the frame moving with velocity $V^{\prime}$. Brodsky
and Primack \cite{Brodsky1969} have shown, using the BSE for atomic hydrogen,
that the Lorentz boost for $P^{\prime}$ mixes the four different Pauli-spinors
$\Omega_{S_{z}}^{S}$ among themselves and the four different Dirac-spinors
$\mathbf{e}_{ij}$ among themselves. The Lorentz boost for a single-particle
fermion in the $z$ direction with velocity $\left\langle \alpha_{z}^{\prime
}\right\rangle =V^{\prime}$ is
\begin{equation}
L=\sqrt{\frac{\gamma^{\prime}+1}{2}}\mathbf{+}\sqrt{\frac{\gamma^{\prime}%
-1}{2}}\alpha_{z}^{\prime}. \label{6.3}%
\end{equation}

For the DV states (\ref{4.40}), with $\Delta_{i}$ given (\ref{4.22a}),
including the plane-wave function $f$, the box-normalized wavefunction in the
rest frame is%
\[
|\Psi,J\rho_{i}\rangle f=|\Psi,J\rho_{i}\rangle e^{-iM_{i}T}/\sqrt{Z_{0}}.
\]
One can boost the DV state with velocity $V^{\prime}\ $so that the total
momentum is $P^{\prime}$. The two-body Lorentz boost $L^{2}$ is the direct
product of the Lorentz boosts $L_{e}\times L_{p}$ such that%
\[
L^{2}=L_{e}\times L_{p}=\left(  \sqrt{\frac{\gamma+1}{2}}\mathbf{+}\sqrt
{\frac{\gamma-1}{2}}\alpha_{ez}\right)  \times\left(  \sqrt{\frac{\gamma+1}%
{2}}\mathbf{+}\sqrt{\frac{\gamma-1}{2}}\alpha_{pz}\right)  ,
\]
where $\gamma$ corresponds to the individual particle boosts $\pm\beta$.
Letting $\sqrt{\gamma^{2}-1}=\beta\gamma$ and using (\ref{5.6}) when operating
on the DV states, one has%
\begin{align}
L^{2}  &  =\frac{\gamma+1}{2}+\frac{\gamma-1}{2}\alpha_{ez}\alpha_{pz}%
+\beta\gamma\frac{(\alpha_{ez}+\alpha_{pz})}{2},\label{6.9b}\\
&  =\gamma\mathbf{+}\beta\gamma\alpha_{z}^{\prime},\nonumber
\end{align}
where%
\begin{equation}
\alpha_{z}^{\prime}=\frac{(\alpha_{ez}+\alpha_{pz})}{2}, \label{6.10}%
\end{equation}
as in (\ref{2.4}). In the frame where $\mathbf{\pi}=\mathbf{0}$, one has
$\mathbf{p}_{e}=\mathbf{p}_{p}$ so that the velocity of the electron and
positron are equal and opposite, $\mathbf{v}_{e}=-\mathbf{v}_{p}$, where
either the electron or the positron is in a negative energy state.
Transforming to either the electron or positron frame, one can add these
velocities relativistically, so that
\[
V^{\prime}=\beta^{\prime}=\pm2\beta/(1+\beta^{2}).
\]
Then the two-body Lorentz boost (\ref{6.9b}) is equivalent to%
\begin{equation}
L^{2}=\sqrt{\frac{\gamma^{\prime}+1}{2}}\mathbf{+}\sqrt{\frac{\gamma^{\prime
}-1}{2}}\alpha_{z}^{\prime}, \label{6.11}%
\end{equation}
where
\[
\gamma^{\prime}=(1-\beta^{^{\prime}2})^{-%
\frac12
}=(1+\beta^{2})/(1-\beta^{2}).
\]
But this is just the boost (\ref{6.3}) with velocity $V^{\prime}=\beta
^{\prime}$ where the rest mass is now $M_{i}^{\prime}$ in (\ref{5.2b}) such
that%
\begin{align}
V^{\prime}  &  =\beta^{\prime}=P^{\prime}/E_{i}^{\prime},\label{6.2}\\
E_{i}^{\prime}  &  =\sqrt{P^{\prime2}+M_{i}^{\prime2}},\ \ \gamma^{\prime
}=E_{i}^{\prime}/M_{i}^{\prime},\ \ \ \beta^{\prime}\gamma^{\prime}=P^{\prime
}/M_{i}^{\prime}.\nonumber
\end{align}
In general, because $\rho_{i}^{\prime}\neq\rho_{i}$, the binding energy has
changed in the moving frame so that $M_{i}^{\prime}\neq M_{i}$ and the
particle cannot be considered a single-particle fermion.

One can now use $L^{2}$ in (\ref{6.11}) to transform the DV states in
(\ref{4.40}) with $\Delta_{i}$ in (\ref{4.22a}) and confirm the results above
in (\ref{5.2b}). Operating with $\alpha_{ez}$ and $\alpha_{pz}$ in
(\ref{6.10}), one finds that
\begin{align}
\alpha_{ez}\Omega_{0}^{0}(\mathbf{e}_{11}-\mathbf{e}_{22}) &  =\alpha
_{pz}\Omega_{0}^{0}(\mathbf{e}_{11}-\mathbf{e}_{22})=-\Omega_{0}%
^{1}(\mathbf{e}_{12}-\mathbf{e}_{21}),\label{6.12}\\
\alpha_{ez}\Omega_{0}^{0}(\mathbf{e}_{12}-\mathbf{e}_{21}) &  =\alpha
_{pz}\Omega_{0}^{0}(\mathbf{e}_{12}-\mathbf{e}_{21})=-\Omega_{0}%
^{1}(\mathbf{e}_{11}-\mathbf{e}_{22}),\nonumber\\
\alpha_{ez}\Omega_{+}^{1}(\mathbf{e}_{11}+\mathbf{e}_{22}) &  =\alpha
_{pz}\Omega_{+}^{1}(\mathbf{e}_{11}+\mathbf{e}_{22})=-\Omega_{-}%
^{1}(\mathbf{e}_{12}+\mathbf{e}_{21}),\nonumber\\
\alpha_{ez}\Omega_{+}^{1}(\mathbf{e}_{12}+\mathbf{e}_{21}) &  =\alpha
_{pz}\Omega_{+}^{1}(\mathbf{e}_{12}+\mathbf{e}_{21})=-\Omega_{-}%
^{1}(\mathbf{e}_{11}+\mathbf{e}_{22}),\nonumber
\end{align}
(and their Hermitian conjugates) so that%
\begin{equation}
\alpha_{z}^{\prime}|\Psi,J\rho_{i}\rangle=\alpha_{ez}|\Psi,J\rho_{i}%
\rangle=\alpha_{pz}|\Psi,J\rho_{i}\rangle.\label{6.10a}%
\end{equation}
The transformed wavefunctions $\Psi_{i}^{\prime}$, after renormalization, are
\begin{align}
\Psi_{i,A}^{0\prime} &  =%
\frac12
\Delta_{i}^{0\prime}\left\{  \sqrt{\gamma^{\prime}+1}\Omega_{0}^{0}%
(\mathbf{e}_{11}-\mathbf{e}_{22})-\sqrt{\gamma^{\prime}-1}\Omega_{0}%
^{1}(\mathbf{e}_{12}-\mathbf{e}_{21})\right\}  /\sqrt{\gamma^{\prime}%
},\label{6.20}\\
\Psi_{i,S}^{0\prime} &  =%
\frac12
\Delta_{i}^{0\prime}\left\{  \sqrt{\gamma^{\prime}+1}\Omega_{0}^{0}%
(\mathbf{e}_{12}-\mathbf{e}_{21})-\sqrt{\gamma^{\prime}-1)}\Omega_{0}%
^{1}(\mathbf{e}_{11}-\mathbf{e}_{22})\right\}  /\sqrt{\gamma^{\prime}%
},\nonumber\\
\Psi_{i,1S}^{1\prime} &  =%
\frac12
\Delta_{i}^{1\prime}\{\Omega_{+}^{1}(\mathbf{e}_{11}+\mathbf{e}_{22}%
)-\Omega_{-}^{1}(\mathbf{e}_{12}+\mathbf{e}_{21})\},\nonumber\\
\Psi_{i,2S}^{1\prime} &  =%
\frac12
\Delta_{i}^{1\prime}\{\Omega_{+}^{1}(\mathbf{e}_{12}+\mathbf{e}_{21}%
)-\Omega_{-}^{1}(\mathbf{e}_{11}+\mathbf{e}_{22})\},\nonumber
\end{align}
where the $\Delta_{i}^{\prime}$ are given by (\ref{4.22b}). The
renormalization factor $\sqrt{1/\gamma^{\prime}}$ comes from the Lorentz
contraction $Z_{0}^{\prime}=Z_{0}/\gamma^{\prime}$ of the transformed plane-wavefunction,%

\begin{equation}
\ f^{\prime}=e^{i(P^{\prime}Z^{\prime}-E_{i}^{\prime}T^{\prime})}/\sqrt
{Z_{0}^{\prime}}.\ \label{6.13}%
\end{equation}
Using (\ref{5.2}) and (6.20), one can readily verify (\ref{5.2b}). One can
also verify directly that the DV states in (\ref{6.20}) are solutions to the
one-body Dirac equation (\ref{2.4}),%
\begin{equation}
(\alpha_{z}^{\prime}P_{Z}^{\prime}+M^{\prime}\gamma_{4}^{2})f^{\prime}%
|\Psi_{i}^{\prime}\rangle=E_{i}^{\prime}f^{\prime}|\Psi_{i}^{\prime}\rangle,
\label{7.8}%
\end{equation}
where $M^{\prime}\gamma_{4}^{2}=\Phi_{C}^{\prime}\alpha_{0}^{2}$ and
$E_{i}^{\prime}=\sqrt{P_{Z}^{\prime}+(M_{i}^{\prime})^{2}}$ as shown in the
Appendix. To confirm that the DV states with $z_{i}=0$ are single-particle
fermions, one has $\rho^{\prime}=\rho$ and $\Phi_{C}^{\prime}=\Phi_{C}$ for
these special DV states, so that $M^{\prime}=M$.

For these `pancake' wavefunctions with $z_{i}=z_{i}^{\prime}=0$, the DV states
for $S_{z}=\pm1$ and $S_{z}=0$ then transform like fermion doublets in which
the Lorentz contraction in $z$ has no effect. Looking more closely at the
spinor states of (\ref{6.20}), one finds that, instead of the expected
single-particle doublet Pauli-spinors $(\chi_{%
\frac12
},\chi_{-%
\frac12
})$, one now has the doublet Dirac-spinors $\left\{  (\mathbf{e}%
_{11}-\mathbf{e}_{22}),(\mathbf{e}_{12}-\mathbf{e}_{21})\right\}  $ for
$\Psi^{0\prime}$ and $\left\{  (\mathbf{e}_{11}+\mathbf{e}_{22}),(\mathbf{e}%
_{12}+\mathbf{e}_{21})\right\}  $ for $\Psi^{1\prime}$, respectively. Also,
one finds that, instead of the expected single-particle Dirac-spinors
$(\mathbf{e}_{1},\mathbf{e}_{2})$, one now has the Pauli-spinors $(\Omega
_{0}^{0},\Omega_{0}^{1})$ for $\Psi^{0\prime}$ and $(\Omega_{+}^{1},\Omega
_{-}^{1})$ for $\Psi^{1\prime}$. For a given DV bound-state doublet, the spin
states $s_{z}(e)$ and $s_{z}(p)$ are completely correlated and act like a
single-particle spin state $s_{z}$. That is, for the Pauli-spinors
$(\Omega_{0}^{0},\Omega_{0}^{1})$, one has $s_{z}(e)=-s_{z}(p)$, whereas for
the Pauli-spinors $(\Omega_{+}^{1},\Omega_{-}^{1})$, one has $s_{z}%
(e)=s_{z}(p)$.

Summarizing, the directed DV doublets with coordinates $r=r_{i}$ and $z_{i}=0$
in the rest frame transform like a single-particle fermion when undergoing a
Lorentz boost because of dynamical symmetry breaking. The rest mass for the
heavy fermions is $M_{i}^{0\prime}=M_{i}^{0}=2e^{2}/r_{i}$ and rest mass for
the light fermions is $M_{i}^{1\prime}=M_{i}^{1}=0$. As shown by (\ref{10.4})
for $z_{i}=0$, the $\Psi^{0\prime}$ DV states must have $J=even$ and the
$\Psi^{1\prime}$ DV states must have $J=odd$. However, for both of these
fermions, the role of Pauli-spinors and Dirac-spinors has been reversed
because either the electron or positron is in a negative energy state. The
$S_{z}=0$ fermions in (\ref{6.20}) have both $S$ and $A$ symmetry under
exchange whereas the $S_{z}=\pm1$ fermions have only $A$ symmetry$.$ Most
importantly, because these DV states are comprised of either $S$ or $A$
symmetry, they are their own antiparticles and the light and heavy DV doublets
are Majorana fermions\textbf{. }

\section{Chirality $\chi$ and Helicity $h$ of the Light Fermions}

The light DV fermions $\Psi_{i}^{1\prime}=\Psi_{i,A}^{1\prime}$ with $z_{i}=0$
in (\ref{6.20}) have $A$ exchange symmetry for $J=odd$ corresponding to the
symmetric delta functions in $\Delta_{i,S}^{1\prime}$ (\ref{4.22b}) and the
$e^{\pm i\phi}$ dependence in (\ref{4.13}). These $\Psi_{i}^{1\prime}$ DV
states have well defined chirality and helicity because the rest mass is
$M_{i}^{1\prime}=0$. The chirality operator $\boldsymbol{\chi}=-\gamma_{5}$
for a single-particle operates on the Dirac-spinors such that, for any
$i=1,2,$%
\[
\boldsymbol{\chi}_{e}\mathbf{e}_{1i}=\mathbf{e}_{2i},\ \boldsymbol{\chi}%
_{p}\mathbf{e}_{i1}=\mathbf{e}_{i2},
\]
and their Hermitian conjugates. When operating on these DV states in
(\ref{6.20}), one can define the chirality operator $\boldsymbol{\chi}%
^{\prime}$ as%
\begin{equation}
\boldsymbol{\chi}^{\prime}=\frac{(\boldsymbol{\chi}_{e}+\boldsymbol{\chi}%
_{p})}{2}, \label{8.1}%
\end{equation}
where $\boldsymbol{\chi}^{\prime}\Psi_{i}^{1\prime}=\boldsymbol{\chi}_{e}%
\Psi_{i}^{1\prime}=\boldsymbol{\chi}_{p}\Psi_{i}^{1\prime}$ .

The helicity operator $\mathbf{h}=\mathbf{\Sigma\cdot}\widehat{\mathbf{P}%
^{\prime}}=\mathbf{\Sigma}_{z}$ for a single-particle operates on the
Pauli-spinors such that%
\begin{align*}
\mathbf{h}_{e}\Omega_{1}^{1}  &  =\Omega_{1}^{1},\ \mathbf{h}_{e}\Omega
_{-1}^{1}=-\Omega_{-1}^{1},\\
\ \mathbf{h}_{p}\Omega_{1}^{1}  &  =\Omega_{1}^{1},\ \mathbf{h}_{p}\Omega
_{-1}^{1}=-\Omega_{-1}^{1}.
\end{align*}
Again, when operating on the DV states (\ref{6.20}), one can define the
helicity operator $\mathbf{h}^{\prime}$ for the light fermions as%
\begin{equation}
\mathbf{h}^{\prime}=\frac{(\mathbf{h}_{e}+\mathbf{h}_{p})}{2}, \label{8.2}%
\end{equation}
where $\mathbf{h}^{\prime}\Psi_{i}^{1\prime}=\mathbf{h}_{e}\Psi_{i}^{1\prime
}=\mathbf{h}_{p}\Psi_{i}^{1\prime}$.

The eigenfunctions of both $\boldsymbol{\chi}^{\prime}$ and $\mathbf{h}%
^{\prime}$ are then%
\begin{align}
\Psi_{i,+}^{1\prime}  &  =(\Psi_{i,1A}^{1\prime}+\Psi_{i,2A}^{1\prime}%
)/\sqrt{2}\label{8.3}\\
&  =%
\frac12
\Delta_{i,S}^{1\prime}\{(\mathbf{e}_{11}+\mathbf{e}_{22}+\mathbf{e}%
_{12}+\mathbf{e}_{21})\Omega_{1}^{1}e^{-i\phi}\},\nonumber\\
\Psi_{i,-}^{1\prime}  &  =(\Psi_{i,1A}^{1\prime}-\Psi_{i,2A}^{1\prime}%
)/\sqrt{2}\nonumber\\
&  =%
\frac12
\Delta_{i,S}^{1\prime}\{(\mathbf{e}_{11}+\mathbf{e}_{22}-\mathbf{e}%
_{12}-\mathbf{e}_{21})\Omega_{-1}^{1}e^{i\phi}\},\nonumber
\end{align}
such that $\boldsymbol{\chi}^{\prime}\Psi_{i,\pm}^{1\prime}=\chi^{\prime}%
\Psi_{i,\pm}^{1\prime}$ and $\mathbf{h}^{\prime}\Psi_{i,\pm}^{1\prime
}=h^{\prime}\Psi_{i,\pm}^{1\prime}$. These states $\Psi_{i,\pm}^{1\prime}$
have chirality $\chi^{\prime}=\pm1$ and helicity $h^{\prime}=\pm1$,
respectively, corresponding to right- and left-handed chirality and to right-
and left-handed helicity. As in the case of a single-particle fermion with no
mass, the chirality and helicity of these light DV Majorana fermions have the
same signs.

\section{Conclusions}

It has been shown that the solutions of the Bethe-Salpeter equation (BSE) for
the atomic bound-states of positronium in (\ref{3.1}) and for the anomalous
bound-states of positronium in (\ref{3.2}), result in two entirely different
forms of positronium. One is the normal atomic form of positronium in which
the electron and positron are bound at atomic distances ($bohr$), and the
other is the anomalous form of positronium in which the particles can be bound
at nuclear distances ($fermi$). Such anomalous bound-states are called
discrete variable (DV) states because they form a bases for the DV
representation in which the relative coordinates $(\rho,\theta)$ between the
electron and positron are quantized at discrete values $(\rho_{i},\theta_{i}%
)$. For the atomic states, the BSE stipulates that the negative energy states
must propagate backward in time with $K_{F}$, whereas, for the DV states, the
BSE stipulates that the negative energy states must propagate forward in time
with $K_{R}$. For the DV states, only bound-state solutions are allowed
because of the time behavior of the negative energy states, and these
bound-states cannot dissociate.

It has also been shown that the properties of the anomalous bound-states and
the atomic bound-states differ radically because, for the former, either the
electron or the positron must be in a negative energy state. It is instructive
to compare and contrast the properties of the atomic and DV bound-state
solutions for positronium, as they are complementary. The atomic bound-states
in the rest frame have low relative momentum $\pi<<m$, and can be treated
using the Coulomb gauge, whereas the DV bound-states in the rest frame have
very high relative momentum $\pi>>m$ and must be treated relativistically with
the Feynman gauge. For the atomic free-states, there is no potential energy
and the momenta are quantized. For the DV bound-states, there is no relative
kinetic energy and the relative coordinates are quantized. The atomic states
are bound mainly by the Coulomb potential whereas the DV bound-states are
bound only by the transverse magnetic potential. The atomic states of
positronium are unstable and decay quickly into photons, whereas the DV states
of positronium are stable and cannot decay. Also, the atomic states can emit
and absorb light, whereas the DV states are dark.

The atomic states of positronium are bosons with total spin $S=1$
(corresponding to the triplets $\Omega_{-1}^{1},\Omega_{0}^{1},\Omega_{1}^{1}%
$) and $S=0$ (corresponding to the singlet $\Omega_{0}^{0}$). The DV states of
positronium are comprised of $S_{z}=0$ doublets $(\Omega_{0}^{0},\Omega
_{0}^{1})$ which have opposite spins $s_{z}(e)=-s_{z}(p)$, and $S_{z}=\pm1$
doublets $(\Omega_{1}^{1},\Omega_{-1}^{1})$ which have aligned spins
$s_{z}(e)=s_{z}(p)$. In both these cases the individual electron and positron
spins are correlated and transform like single-particle fermions by a Lorentz
boost when they are in the plane perpendicular to the direction of motion.

The fermion nature of the DV states occurs because the wavefunctions are
Lorentz contracted in the direction of motion $\widehat{z}$ resulting in the
spherical symmetry being reduced to cylindrical symmetry. The relative
coordinates $(r,z)$ can then be quantized in the $z=0$ plane with $r=r_{i}$.
For such states the binding energy is independent of the motion and the DV
doublet states behave like single-particle fermions. The $S_{z}=0$ fermions
for $(\Omega_{0}^{0},\Omega_{0}^{1})$ are heavy particles with mass $M_{i}%
^{0}=2e^{2}/r_{i}$ and the $S_{z}=\pm1$ fermions for $(\Omega_{-1}^{1}%
,\Omega_{1}^{1})$ are light particles with mass $M_{i}^{1}=0$. Because the DV
fermions are either symmetric $\Psi_{S}=\sqrt{%
\frac12
}(\Psi_{+-}+\Psi_{-+})$ or antisymmetric $\Psi_{A}=\sqrt{%
\frac12
}(\Psi_{+-}-\Psi_{-+})$, depending on their angular momentum $J$, they are
their own antiparticle and are therefore Majorana fermions. However, for the
light fermions, which can only occur in the $z=0$ plane, only the $\Psi_{A}$
states are possible. They also have well defined helicity and chirality
because their mass $M_{i}$ is zero. Thus, the solutions of the BSE equation
for positronium result in two forms of matter with contrasting but
complementary characteristics.

It does not appear that the DV states are relevant, per se, to atomic physics.
The question arises as to whether there are any particles which have the
properties of the DV bound-state fermions. It appears that the properties of
the DV fermions are consistent with those of light and heavy neutrinos. One
can then hypothesize that the DV states corresponding to the $S_{z}=\pm1$
light fermions are electron neutrinos and the DV states corresponding to the
$S_{z}=0$ heavy fermions are `sterile' neutrinos. The `sterile' type of
neutrino has been surmised by some particle physicists as an explanation for
dark matter \cite{Boyarsky2019}. The sterile neutrino would then have a mass
of $\sim MeV$ if the electron and positron were bound at nuclear distances
($fermi$). The fact that these DV bound-states are Majorana fermions and are
not single point particles would then explain the violation of parity and
charge conjugation by the weak force. These light Majorana fermions can have
either left- or right-handed helicity (with the same chirality) which is then,
presumably, selected by the weak force to give the observed handedness of neutrinos.

The fact that these light and heavy Majorana fermions are dark, stable, and
are composed of equal amounts of matter (electrons) and antimatter (positrons)
could then simultaneously explain both the apparent absence of antimatter in
the universe as well as the apparent presence of dark matter. Further
investigation of this hypothesis is necessary to show that the light fermions
have other properties of the electron neutrinos besides the ones shown here.
One needs to show that the light fermions have the measured weak force
cross-sections and are consistent with the proven theory of vector bosons.

\section*{Acknowledgments}

The author is grateful to Dr. Tony Scott for his continued support and
invaluable discussions over the past several years. Scott's work with Drs.
Shertzer and Moore led to the author's interest in the TBDE and subsequent
discovery of the DV states. The author thanks Dr. Max Standage for his
interest and help along with Drs. Robert Sang and Howard Wiseman for their
support while at Griffith University. The author is grateful to Dr. Brian
Kendrick for suggesting the use of the discrete variable representation and
Drs. James Colgan, Arthur Voter, Peter Milonni and Joel Kress for their
hospitality and help while at Los Alamos National Laboratory. Dr. Gordon Drake
suggested the separability of the atomic and anomalous states which led to the
extension of the Bethe-Salpeter equation for DV states. Finally, Dr. George
Csanak has made many useful comments during the revision of this paper and Dr.
William Harter has been steadfast in his support over many years.

\appendix

\section{Lorentz Boost of Operators Using Adjoint and Conjugate Spinors}

Operator equations using the Lorentz boost $L^{2}$ involve difficulties
because $L^{2}$ is not a unitary transform. The Lorentz boosts of wavefunction
$\Psi$ and operator $O$ comprised of Dirac spinors have the properties%
\begin{align}
\Psi^{\prime}  &  =L^{2}\Psi,\ O^{\prime}=L^{2}OL^{-2},\qquad\label{12.0}\\
L^{-2}  &  =\gamma_{4}^{2}L^{2}\gamma_{4}^{2},\ \ \ \gamma_{4}^{2}=\gamma
_{e4}\gamma_{p4},\ \nonumber
\end{align}
where $L^{-2}$ is the inverse transform. Define the adjoint spinor to be
\begin{align}
\langle\overline{\Psi}|  &  =\left\langle \Psi\right\vert \gamma_{4}%
^{2},\label{12.1}\\
\langle\overline{\Psi^{\prime}}|  &  =\left\langle \Psi^{\prime}\right\vert
\gamma_{4}^{2}=\left\langle \Psi\right\vert L^{2}\gamma_{4}^{2}=\left\langle
\Psi\right\vert \gamma_{4}^{2}\gamma_{4}^{2}L^{2}\gamma_{4}^{2}=\left\langle
\overline{\Psi}\right\vert L^{-2},\nonumber
\end{align}
where $\left\langle \Psi\right\vert $ is the conjugate spinor. There are two
different expressions for the expectation value of an operator $O$ which is
Lorentz boosted. Define the adjoint expectation of operator $O$ to be
$\langle\overline{\Psi}\left\vert O\right\vert \Psi\rangle$ and the conjugate
expectation to be $\left\langle \Psi\left\vert O\right\vert \Psi\right\rangle
$. The adjoint expectation of $O$ is a Lorentz invariant,
\begin{equation}
\langle\overline{\Psi^{\prime}}\left\vert O^{\prime}\right\vert \Psi^{\prime
}\rangle=\left\langle \overline{\Psi}\left\vert O\right\vert \Psi\right\rangle
\label{12.2}%
\end{equation}
which for $O=1$ corresponds to the Lorentz scalar,
\[
\langle\overline{\Psi^{\prime}}|\Psi^{\prime}\rangle=\langle\overline{\Psi
}|\Psi\rangle.
\]
This invariant form of the expectation should be used when transforming to the
moving frame. For example, using (\ref{12.2}) for the kinetic operator $K$,
one finds that, for the DV states,%
\begin{equation}
\langle\overline{\Psi^{\prime}}\left\vert K^{\prime}\right\vert \Psi^{\prime
}\rangle=\left\langle \overline{\Psi}\left\vert K\right\vert \Psi\right\rangle
=0, \label{12.3}%
\end{equation}
so that the transformed wavefunctions in (\ref{4.40}) are still anomalous
states with zero kinetic energy.

The DV potential can be written%

\[
\Phi_{DV}=-\frac{e^{2}}{\rho}(1-\boldsymbol{\alpha}_{e}\cdot\boldsymbol{\alpha
}_{p})=\Phi_{C}\alpha_{0}^{2},
\]
where $\Phi_{C}$ is the Coulomb potential. The operators $\alpha_{0}^{2}$ and
$\gamma_{4}^{2}$ commute so that $\alpha_{0}^{2}\gamma_{4}^{2}=\gamma_{4}%
^{2}\alpha_{0}^{2}$. To find the transform properties of the DV potential, one
must consider the Lorentz invariant operator $\gamma_{0}^{2}=\gamma_{eu}%
\gamma_{pu}$, which is the scalar product of the two four-vectors $\gamma
_{eu}$ and $\gamma_{pu}$. This scalar operator transforms such that
\begin{equation}
\gamma_{0}^{\prime2}=\gamma_{eu}^{\prime}\gamma_{pu}^{\prime}=\gamma
_{eu}\gamma_{pu}=\gamma_{0}^{2}, \label{12.4}%
\end{equation}
or equivalently,%
\begin{equation}
\gamma_{0}^{\prime2}=\gamma_{4}^{\prime2}\alpha_{0}^{\prime2}=\gamma_{4}%
^{2}\alpha_{0}^{2}=\gamma_{0}^{2}. \label{12.4a}%
\end{equation}
Unlike the operator $\gamma_{0}^{2}$, the operator $\alpha_{0}^{2}$ is not a
Lorentz invariant as seen in (\ref{12.4a}). From the above equations one finds
that
\begin{align*}
\left\langle \overline{\Psi}\left\vert \gamma_{0}^{2}\right\vert
\Psi\right\rangle  &  =\left\langle \overline{\Psi}\left\vert \gamma_{4}%
^{2}\alpha_{0}^{2}\right\vert \Psi\right\rangle =\left\langle \Psi\left\vert
\alpha_{0}^{2}\right\vert \Psi\right\rangle ,\\
\langle\overline{\Psi^{\prime}}\left\vert \gamma_{0}^{\prime2}\right\vert
\Psi^{\prime}\rangle &  =\langle\overline{\Psi^{\prime}}\left\vert \gamma
_{4}^{2}\alpha_{0}^{2}\right\vert \Psi^{\prime}\rangle=\left\langle
\Psi^{\prime}\left\vert \alpha_{0}^{2}\right\vert \Psi^{\prime}\right\rangle .
\end{align*}
Letting $\left\langle \overline{\Psi}\left\vert \gamma_{0}^{2}\right\vert
\Psi\right\rangle =\langle\overline{\Psi^{\prime}}\left\vert \gamma
_{0}^{\prime2}\right\vert \Psi^{\prime}\rangle$, one finds the useful result%
\begin{equation}
\left\langle \Psi\left\vert \alpha_{0}^{2}\right\vert \Psi\right\rangle
=\left\langle \Psi^{\prime}\left\vert \alpha_{0}^{2}\right\vert \Psi^{\prime
}\right\rangle . \label{12.5}%
\end{equation}

The instantaneous Lorentz potential $\Phi_{L}$ is given by%
\[
\Phi_{L}=\Phi_{C}\gamma_{0}^{2}=\gamma_{4}^{2}\Phi_{DV}.
\]
This potential $\Phi_{L}$ can be derived in the momentum representation from
the invariant potential used by Salpeter \cite{Salpeter1952},
\begin{align*}
G(k_{u})  &  =-e^{2}(\gamma_{0}^{2}/k_{0}^{2}),\\
k_{0}^{2}  &  =\mathbf{k\cdot k-\varpi}^{2}.
\end{align*}
For the instantaneous Lorentz potential in the momentum representation, let
$\mathbf{\varpi}^{2}=0$. Now let the two-body mass operator for the DV states
be defined by%
\begin{equation}
M\gamma_{4}^{2}=\Phi_{L}\gamma_{4}^{2}=\Phi_{C}\alpha_{0}^{2}=\Phi_{DV},
\label{12.6}%
\end{equation}
so that the two-body mass operator is the analogy of the one-body mass
operator $m\gamma_{4}.$ One must then use the two-body mass operator,
\begin{equation}
M^{\prime}\gamma_{4}^{2}=\Phi_{L}^{\prime}\gamma_{4}^{2}=\Phi_{C}^{\prime
}\alpha_{0}^{2}, \label{12.7}%
\end{equation}
when finding the mass $M^{\prime}$ in the moving frame in analogy with the
one-body case.

From (\ref{12.6}) and (\ref{12.7}), one finds that the adjoint expectation of
the Lorentz potential $\Phi_{L}$ and $\Phi_{L}^{\prime}$ is
\begin{align}
M  &  =\left\langle \overline{\Psi}\left\vert \Phi_{L}\right\vert
\Psi\right\rangle =\left\langle \Psi|\Phi_{C}\alpha_{0}^{2}|\Psi\right\rangle
,\label{12.8}\\
M^{\prime}  &  =\langle\overline{\Psi^{\prime}}\left\vert \Phi_{L}^{\prime
}\right\vert \Psi^{\prime}\rangle=\left\langle \Psi^{\prime}|\Phi_{C}^{\prime
}\alpha_{0}^{2}|\Psi^{\prime}\right\rangle .\nonumber
\end{align}
The Lorentz potential is a Lorentz invariant when $\rho^{\prime}=\rho$ or when
$\Phi_{C}^{\prime}=\Phi_{C}$. Combining the result (\ref{12.5}) with the
Lorentz contraction of $\rho$ to $\rho^{\prime}$ in the moving frame allows
one to find $M$ and $M^{\prime}$ in (\ref{12.8}). Using $M$ in (\ref{12.8})
and (\ref{10.3a}) confirms (\ref{5.2}). Using $M^{\prime}$ in (\ref{12.8}) and
(\ref{10.3b}) confirms (\ref{5.2b}), where
\begin{align}
M_{i}^{0\prime}  &  =\langle\Psi_{i}^{0\prime}|\Phi_{C}^{\prime}\alpha_{0}%
^{2}|\Psi_{i}^{0\prime}\rangle=2e^{2}/\rho_{i}^{\prime},\label{12.9}\\
M_{i}^{1\prime}  &  =\langle\Psi_{i}^{1\prime}|\Phi_{C}^{\prime}\alpha_{0}%
^{2}|\Psi_{i}^{1\prime}\rangle=0.\nonumber
\end{align}
This results in the two-body equation for the DV states in the moving frame%
\begin{align}
(\alpha_{z}^{\prime}P_{Z}^{\prime}+M^{\prime}\gamma_{4}^{2})f^{\prime}%
|\Psi_{i}^{\prime}\rangle &  =E_{i}^{\prime}f^{\prime}|\Psi_{i}^{\prime
}\rangle,\label{12.10}\\
\ f^{\prime}  &  =e^{i(P^{\prime}Z^{\prime}-E_{i}^{\prime}T^{\prime})}%
/\sqrt{Z_{0}^{\prime}}.\nonumber
\end{align}
as in (\ref{7.8}). This equation is equivalent to the single-body Dirac
equation for a fermion only if $M^{\prime}=M$ such that the rest mass is
constant in the moving frame.

\bigskip
\bibliographystyle{apsrev}
\bibliography{chrisref}

\end{document}